\documentclass[a4paper,11pt]{article}

\usepackage{braket}
\usepackage{amsmath}
\usepackage{amssymb}
\usepackage{color}
\usepackage{cite}
\usepackage{hyperref}
\usepackage{graphicx}
\usepackage{subfig}

\makeatletter
\@addtoreset{equation}{section}
\renewcommand{\theequation}{\thesection.\@arabic\c@equation}
\makeatother
\makeatletter
\renewcommand\appendix{\par
  \setcounter{section}{0}%
  \setcounter{subsection}{0}%
  \gdef\thesection{Appendix \@Alph\c@section }
  \renewcommand{\theequation}
  {\Alph{section}.\arabic{equation}}
}
\makeatother

\def \be {\begin{equation}}
\def \ee {\end{equation}}
\def \ba {\begin{array}}
\def \ea {\end{array}}
\def \bea{\begin{eqnarray}}
\def \eea{\end{eqnarray}}

\newcommand{\f}{\frac}
\newcommand{\s}{\sqrt}
\newcommand{\ep}{\epsilon}

\def \s {\sigma}

\def \p {\partial}
\def \f {\frac}

\def \nn {\nonumber}

\def \la {\leftarrow}

\def \hs {\hspace}

\def \inf {\infty}

 \def\la{\langle}
 \def\lb{\rangle}
 \def\ep{\epsilon}

\title{\textbf{Entanglement Entropy for Descendent Local Operators in 2D CFTs}}
\author{
Bin Chen$^{1,2,3}$\footnote{bchen01@pku.edu.cn}\,
Wu-Zhong Guo$^{4,5}$\footnote{wuzhong@itp.ac.cn}\,
Song He$^{5,4}$\footnote{hesong17@gmail.com}\, and
Jie-qiang Wu$^{1}$\footnote{jieqiangwu@pku.edu.cn}}
\date{}

\begin{document}

\maketitle

\begin{center}
{\it
$^{1}$Department of Physics and State Key Laboratory of Nuclear Physics and Technology, Peking University, Beijing 100871, P.R.\! China\\
\vspace{2mm}
$^{2}$Collaborative Innovation Center of Quantum Matter,  \\Beijing 100871, P.~R.~China\\
$^{3}$Center for High Energy Physics, Peking University,  \\Beijing 100871, P.~R.~China\\
$^{4}$State Key Laboratory of Theoretical Physics,
Institute of Theoretical Physics, Chinese Academy of Science,\\
Beijing 100190, P. R. China\\
$^{5}$Yukawa Institute for Theoretical Physics, Kyoto University, Kitashirakawa Oiwakecho,\\
Sakyo-ku, Kyoto 606-8502, Japan\\
}
\vspace{10mm}
\end{center}

\begin{abstract}
We mainly study the R\'enyi entropy and entanglement entropy of the states locally excited by the descendent operators in two dimensional conformal field theories (CFTs). In rational CFTs, we prove that the increase of entanglement entropy and R\'enyi entropy for a class of descendent operators, which are generated by
$\cal{L}^{(-)}\bar{\cal{L}}^{(-)}$ onto the primary operator, always coincide with the logarithmic of quantum dimension of the corresponding primary operator. That means the R\'enyi entropy and entanglement entropy for these descendent operators are the same as the ones of their corresponding primary operator. For 2D rational CFTs with a boundary, we confirm that the R\'enyi entropy always coincides with the logarithmic of quantum dimension of the primary operator during some periods of the evolution.
Furthermore, we consider more general descendent operators generated by $\sum_{} d_{\{n_i\}\{n_j\}}(\prod_{i} L_{-n_i}\prod_{j}{\bar L}_{-n_j})$ on the primary operator. For these operators,  the entanglement entropy and R\'enyi entropy get additional  corrections, as the mixing of holomorphic and anti-holomorphic Virasoro generators enhance the entanglement. Finally, we employ perturbative CFT techniques to evaluate the R\'enyi entropy of the excited operators in deformed CFT. The  R\'enyi and entanglement entropies are increased, and get  contributions not only from local excited operators but  also from global deformation of the theory.
\end{abstract}

\baselineskip 18pt
\thispagestyle{empty}

\newpage

\section{Introduction}
The entanglement entropy (EE) and the entanglement R\'enyi entropy (RE) are very helpful quantities to study global or non-local structures in quantum field theories.  For example, computing  topological  entanglement entropy can characterize the topological order of the system \cite{wen}. In \cite{FFN}  a relationship between the topological entanglement entropy and boundary entropy has been pointed out. And in \cite{CC}  the relationship between the boundary entropy and entanglement entropy has been explored.  Both the entanglement and R\'enyi entropes of a subsystem $A$ are  defined with respect to the reduced density matrix $\rho_A$, which is obtained by tracing out the degrees of freedom in the complement of $A$ from the original density matrix $\rho$. Formally, the $n$-th R\'enyi entanglement entropy is defined by $S^{(n)}_A=\log\mbox{Tr}[\rho_A^n]/(1-n)$.
If  the limit $n\to 1$ is well-defined,  $\displaystyle{\lim_{n\to 1}}S^{(n)}_A$ coincides with the von-Neumann entropy exactly,  which defines the entanglement entropy. This leads to standard replica trick  to in compute  the entanglement entropy in the quantum field theory.

There could be  topological contribution to the entanglement entropy even for gapless theories, e.g. conformal field theories (CFTs). In 2D rational CFTs, it was found\cite{He:2014mwa} that  for the locally primary excited states,  the R\'enyi entropy difference is related to the quantum dimension \cite{Moore:1988uz}\cite{Verlinde:1988sn} of the primary operator, which is a kind of topological quantity. The computations of the entanglement entropies for locally excited states have been formulated in
\cite{Alcaraz:2011tn} \cite{Palmai:2014jqa}\cite{Nozaki:2014hna} in the field theory side. The entanglement entropy
for local free scalar  have been investigated in \cite{Nozaki:2014hna}\cite{Nozaki:2014uaa} (see also \cite{Shiba:2014uia} for the operator approach).
 In the large $N$ CFTs , the entanglement
entropies for locally excited states have been discussed in \cite{Caputa:2014vaa,Asplund:2014coa,Caputa:2014eta}. In a more recent paper \cite{Guo:2015uwa}, the R\'enyi entropy difference $ \Delta S^{(n)}_A$ of the local excited states in a boundary conformal field theory (BCFT) has been studied. In \cite{Das:2015oha} the left-right entanglement entropy for a boundary state has been discussed.

 In all the study of quantum entanglement of local operator, only the primary operators have been discussed.\footnote{There has been some discussion on the entanglement entropy for descendant excited states in \cite{Palmai:2014jqa}.} It is interesting to investigate the  similar effect from a local descendent operators. Among the descendants, the stress tensor is of particular interest, as it is related to the energy density in the field theory, and is dual to a graviton in the holographic dual. In this paper, we generalize the previous study \cite{He:2014mwa}\cite{Guo:2015uwa} on the R\'enyi entropy difference to the descendent operators with or without a boundary condition. First of all, we compute $\Delta S^{(2)}$ explicitly for some simple descendent operators  in minimal models,  and we find that in these cases the entropy difference is exactly  the logarithmic of the quantum dimension of corresponding primary operator. Moreover, we study the descendant operators in a general theory.  The picture is that if the local operator is of the form $ \cal{L}^{(-)}\bar{\cal{L}}^{(-)} O$, the entropy difference is the same as the one for the local primary operator.  This could be explained by
studying the divergent terms in the correlation function.  However for a more generic operator like
\be V=\sum_{\{n_i\},\{\bar{n}_j\}}d_{\{n_i\},\{\bar{n}_j\}}\prod_{i,j}L_{-n_i}\bar L_{-n_j}O(w,\bar{w})\label{first-eq} \ee
we find that  the R\'enyi entropies and entanglement entropy  have additional contribution. Such additional correction originates from the entanglement in the excited states. We give a systematic recipe to calculate the entanglement entropy for general descendant operators. 

We also investigate the R\'enyi entropy for the states excited by a local descendent operator in the rational CFTs with a boundary. The boundaries introduced here do not break the conformal symmetry. Such kinds of theories are called the boundary conformal field theories (BCFT). We find that the R\'enyi entropies excited by the descendants in BCFTs are similar to the one discussed in \cite{He:2014mwa}\cite{Guo:2015uwa}. If the descendent states are generated by the operator (1.1), the boundary does change the time evolution of the R\'enyi entropy but it does not change the maximal value of the R\'enyi entropy.  Finally, we investigate the R\'enyi entropies in a deformed CFT perturbatively.  If the deformation can be seen as a perturbation, the R\'enyi entropies of locally excited states are just the summation of contributions from the local excitation and the one from global deformation.



The layout of this paper is as follows. In section 2, we would like to study the R\'enyi entropy of the local descendent state with or without a boundary. In section 3, we move to study the R\'enyi entropy of the local excited states in 2D CFT with deformation by chemical potentials. In these deformed theories, we obtain the R\'enyi entropy of a subsystem with time evolution. In section 4, we devote to the conclusions and discussions. In appendix, we  list some technics which are very useful in our analysis.

While proceeding with this project, we noticed a recent papers \cite{Pawel} appearing in arXiv, which has overlap with our discussion  in the sections 2.3 and 2.4.

\section{R\'enyi entropy of descendent operator}

\subsection{Setup in 2D CFT}

 Consider an excited state which is defined by acting a primary or a descendent operator $V_a$ on the vacuum $|0\lb$ in a two dimensional CFT. We make use of the Euclidean formulation and introduce the complex coordinate $(w,\bar{w})=(x+i\tau,x-i\tau)$ on $R^2$ such that $\tau$ and $x$ are the Euclidean time and the space respectively. We insert the operator ${V}_a$ at $x=-l<0$ and consider its real time-evolution from time $0$ to $t$ under the Hamiltonian $H$. The corresponding density matrix can be expressed as following:
\bea
\rho(t)&=&{\mathcal N}\cdot e^{-iHt}e^{-\ep H}{V}_a(-l)|0\lb\la 0|{{V}}^{\dagger}_a(-l)e^{-\ep H}e^{iHt} \nonumber\\
&=& {\mathcal N}\cdot{V}_a(w_2,\bar{w_2})|0\lb\la 0|
{{V}}^{\dagger}_a(w_1,\bar{w}_1),
\eea
where ${\mathcal N}$ is fixed by requiring Tr$\rho(t)=1$, and
\bea
&& w_1=i(\epsilon -it)-l, \ \ w_2 = -i(\epsilon+it)-l,   \label{wco} \\
&& \bar{w}_1=-i(\ep-it)-l,\ \ \bar{w}_2=i(\epsilon+it)-l.  \label{wcor}
\eea
The infinitesimal positive parameter $\ep$ is an ultraviolet regularization factor.  We treat
$\ep\pm it$ as purely imaginary numbers until the end of the calculations, as in \cite{Calabrese:2005in,Nozaki:2014hna,Nozaki:2014uaa}.

To calculate $\Delta S^{(n)}_A$, we can apply the replica method in the path-integral formalism
by generalizing the formulation for the ground state \cite{CC} to our excited states \cite{Nozaki:2014hna}. It leads to a $n$-sheeted Riemann surface $\Sigma_n$ with $2n$ operators $V_a$ inserted. We choose the subsystem $A$ to be an interval $0\leq x\leq L$ at $\tau=0$. In section 2.2, 2.3, we would like to take $L\rightarrow \infty$ as a warm up. For later parts, we will consider the situation with finite $L$.

Finally, the $\Delta S_A^{(n)}$ can be computed as
\bea
\Delta S_A^{(n)}
&&=\!\f{1}{1-n} \Biggl[\log{\left\langle{{V}}^{\dagger}_a(w_1,\bar{w}_1){V}_a(w_2,\bar{w}_2)
 \cdots {V}_a(w_{2n},\bar{w}_{2n})\right\rangle_{\Sigma_n}}\nonumber\\
&& \ \ \ \ -n\log\left\langle{{V}}^{\dagger}_a(w_1,\bar{w}_1)
{V}_a(w_2,\bar{w}_2)\right\rangle_{\Sigma_1}\Biggr], \label{replica}
\eea
where $(w_{2k+1},w_{2k+2})$ for $k=1,2,...,n-1$ are $n-1$ replicas of $(w_1,w_2)$ in the $k$-th sheet of $\Sigma_n$. The term in the first line is given by a $2n$-point correlation function on $\Sigma_n$, while the one in the second line is given by $n$ two-point functions on $\Sigma_1$. Here $\Delta_a$ is the (chiral and anti-chiral) conformal dimension of the operator ${O}_a$.

\begin{figure}
  \centering
  \includegraphics[width=5cm]{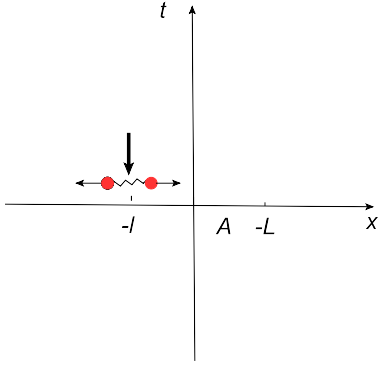}\\
  \caption{This figure is to show our setup in two dimensional plane $\omega=x+i t$ . The system will be triggered at $x=-l$ and there are left- and right-moving quasi-particles at $t=0$.}\label{fig1}
\end{figure}

\subsection{Convention}

Let us study the second R\'enyi entanglement entropy in details to set up the convention.  The calculations of
$\Delta S^{(2)}_A$ is related to the four-point functions which we know pretty well for exactly solvable CFTs.

To compute $\Delta S^{(2)}_A$ we need to introduce the coordinate $z_i$, which is related to $w_i$ by the conformal map $w_i=z^2_i$.
First the coordinates $w_i$ and $z_i$ behave like
\bea
&& w_1=i\epsilon+t-l\equiv re^{i\theta_1}=(z_1)^2, \nonumber\\
&& w_2=-i\epsilon+t-l\equiv se^{i\theta_2}=(z_2)^2, \nonumber\\
&& w_3=(i\epsilon+t-l)e^{2\pi i}\equiv re^{i(2\pi+\theta_1)}=(z_3)^2, \nonumber\\
&& w_4=(-i\epsilon+t-l)e^{2\pi i}\equiv se^{i(2\pi+\theta_2)}=(z_4)^2. \label{wcoordinate}
\eea
Thus we find
\bea
&& z_1=-z_3=\sqrt{w_1}=\sqrt{r}e^{i\theta_1/2}=i\sqrt{l-t-i\epsilon}, \nonumber\\
&& z_2=-z_4=\sqrt{w_2}=\sqrt{s}e^{i\theta_2/2}=i\sqrt{l-t+i\epsilon}. \label{zcor}
\eea
The most important point is that we should treat $\pm i\epsilon+t$ as a pure imaginary number in all algebraic calculations and take $t$ to be real only in the final expression of the entropy.
This is a standard prescription to make analytical continuation of Euclidean theory into its Lorentzian version. Therefore we should identify
\bea
(r\cos\theta_1,r\sin\theta_1)=(-l,\epsilon-it), \hs{3ex} (s\cos\theta_2,s\sin\theta_2)=(-l,-\epsilon-it).
\eea
This leads to
\bea
&& r=\sqrt{l^2+(\epsilon-it)^2},\ \ s=\sqrt{l^2+(-\epsilon-it)^2}, \nonumber\\
&& rs=\sqrt{(l^2+\epsilon^2-t^2)^2+4\epsilon^2t^2},\ \ r^2+s^2=2(l^2+\epsilon^2-t^2), \nonumber\\
&& \cos(\theta_1-\theta_2)=2\cos^2\left(\frac{\theta_1-\theta_2}{2}\right)-1=
\frac{l^2-\epsilon^2-t^2}{\sqrt{(l^2+\epsilon^2-t^2)^2+4\epsilon^2t^2}}. \label{rs}
\eea

To evaluate the four point-correlation functions, it is useful to focus on the ratio
\bea
&& z=\frac{z_{12}z_{34}}{z_{13}z_{24}}=\frac{-(l-t)+\sqrt{(l-t)^2+\epsilon^2}}{2\sqrt{(l-t)^2+\epsilon^2}},\nonumber\\
&& \bar{z}=\frac{\bar{z}_{12}\bar{z}_{34}}{\bar{z}_{13}\bar{z}_{24}}
=\frac{-(l+t)+\sqrt{(l+t)^2+\epsilon^2}}{2\sqrt{(l+t)^2+\epsilon^2}}, \label{zij}
\eea
where $z_{ij}=z_i-z_j$. Note that there is a useful relation
\bea
1-z=\f{z_{14}z_{23}}{z_{13}z_{24}}.
\eea

We are interested in the two limits (i) $l>>t>>\epsilon$ (early time) and (ii) $t>>l>>\epsilon$ (late time). From (\ref{zij}) we find that they correspond to
\bea
&& (i)\ \  z\simeq \bar{z}\simeq \frac{\epsilon^2}{4l^2}\ (\to 0),\nonumber\\
&& (ii)\ \  z\simeq 1-\frac{\epsilon^2}{4t^2}\ (\to 1), \ \ \ \ \bar{z}\simeq \frac{\epsilon^2}{4t^2}\ (\to 0).\label{earlytimelate}
\eea
Note that the second limit is non-trivial in that it does not respect the complex conjugate relation because of our analytical continuation of $t$.

\subsection{Second Renyi entanglement entropy in terms of 4-pt functions}

Let us first consider the R\'enyi entropy of the simplest descendent operator $L_{-1}O$, where $O$ is a primary field.
The four-point function on $n$-sheeted Riemann surface $\Sigma_n$ can be mapped to the one on $R^2$ by the conformal map
$w=z^n$.  For the second R\'enyi entropy,  we find
\bea\label{four}
&& \langle L_{-1}O(w_1,\bar{w}_1)L_{-1}O(w_2,\bar{w}_2)L_{-1}O(w_3,\bar{w}_3)L_{-1}O(w_4,\bar{w}_4)\rangle_{\Sigma_2} \nonumber\\
& =&\prod_{i=1}^{4}\left|\f{dw_i}{dz_i}\right|^{-2\Delta}\Bigg( \langle  \left(-\Delta{\partial{z_1}\over \partial{w_1}}{\partial^2 w_1\over \partial z_{1}^2}+{\partial z_1 \over \partial w_1}\partial_{z_1}\right) O(z_1,\bar{z}_1)  \left(-\Delta{\partial{z_2}\over \partial{w_2}}{\partial^2 w_2\over \partial z_{2}^2}+{\partial z_2 \over \partial w_2}\partial_{z_2}\right)O(z_2,\bar{z}_2) \nonumber\\&&\left(-\Delta{\partial{z_3}\over \partial{w_3}}{\partial^2 w_3\over \partial z_{3}^2}+{\partial z_3 \over \partial w_3}\partial_{z_3}\right) O(z_3,\bar{z}_3) \left(-\Delta{\partial{z_4}\over \partial{w_4}}{\partial^2 w_4\over \partial z_{4}^2}+{\partial z_4 \over \partial w_4}\partial_{z_4}\right)O(z_4,\bar{z}_4)\rangle_{\Sigma_1}\Bigg)\nonumber\\
&& =2^{-8\Delta}|z_1z_2z_3z_4|^{-2\Delta}\cdot  N_1 N_2 N_3 N_4 \Bigg(\Delta^4\langle O(z_1,\bar{z}_1)O(z_2,\bar{z}_2)O(z_3,\bar{z}_3)O(z_4,\bar{z}_4)\rangle_{\Sigma_1}\nonumber\\&&-\Delta^3\sum_{i=1}^4 {M_i \over N_i}\partial_{z_i}\langle O(z_1,\bar{z}_1)O(z_2,\bar{z}_2)O(z_3,\bar{z}_3)O(z_4,\bar{z}_4)\rangle_{\Sigma_1}\nonumber\\&&+
\Delta^2 \sum_{i\neq j}^4 {M_i M_j\over N_i N_j}\partial_{z_i}\partial_{z_j}\langle O(z_1,\bar{z}_1)O(z_2,\bar{z}_2)O(z_3,\bar{z}_3)O(z_4,\bar{z}_4)\rangle_{\Sigma_1} \nonumber\\
&& -\Delta \sum_{ i\neq j\neq k}^4 {M_i M_j M_k\over N_i N_j N_k}\partial_{z_i}\partial_{z_j}\partial_{z_k}\langle O(z_1,\bar{z}_1)O(z_2,\bar{z}_2)O(z_3,\bar{z}_3)O(z_4,\bar{z}_4)\rangle_{\Sigma_1}+\nonumber\\
&& \frac{M_1M_2 M_3 M_4}{N_1N_2N_3N_4}\partial_{z_1}\partial_{z_2}\partial_{z_3}\partial_{z_4}\langle O(z_1,\bar{z}_1)O(z_2,\bar{z}_2)O(z_3,\bar{z}_3)O(z_4,\bar{z}_4)\rangle_{\Sigma_1}\Bigg)
\eea
where $\Delta$ is the chiral conformal dimension of the primary operator $O$, and
\be
N_i={\partial{z_i}\over \partial{w_i}}{\partial^2 w_i\over \partial z_{i}^2},\hspace{3ex} M_i={\partial z_i \over \partial w_i}.
 \ee
 Since $N_i$ and $M_i$ do not contribute to the divergent terms in both the late time limit and the early time limit, we can take them as a factor to simplify our analysis. In (\ref{four}),  the last term in the summation diverge fastest in the small $\epsilon$ limit, so we only keep this term in the later calculation.
Note also that $rs$ behaves  in the two limits as
\bea
&&(i) \mbox{Early time}:\  rs\simeq l^2,\nonumber\\
&& (ii) \mbox{Late time}:\  rs\simeq t^2.
\eea
Due to the conformal symmetry, the four-point function on $R^2$ can be expressed as
\be
\langle O(z_1,\bar{z}_1)O(z_2,\bar{z}_2)O(z_3,\bar{z}_3)O(z_4,\bar{z}_4)\rangle_{\Sigma_1}
=|z_{13}z_{24}|^{-4\Delta}\cdot G(z,\bar{z}), \label{gf}
\ee
where $(z,\bar{z})$  are given by (\ref{zij}).

The two-point function is
\bea
\langle L_{-1}O(w_1,\bar{w}_1)L_{-1}O(w_2,\bar{w}_2)\rangle_{\Sigma_1}
=\partial_{w_1}\partial_{w_2}\f{{\mathcal{N}}}{|w_{12}|^{4\Delta}}=\frac{2\Delta(2\Delta+1){\mathcal{N}}}{(2\epsilon)^{4\Delta+2}},
\eea
where ${\mathcal{N}}$ is the normalization of the primary operator $O$.
 Note that the four-point function is proportional to
${\mathcal{N}}^2$ and the $\Delta S^{(2)}_A$ is of course independent of ${\mathcal{N}}$.

In the late time limit (ii), we finally find that the ratio $Z_2/(Z_1)^2$ in (\ref{replica}) is expressed in terms of the four-point function on $R^2$:
\bea
&& \mbox{Tr}\rho_A^2=\f{\langle L_{-1}O(w_1,\bar{w}_1)L_{-1}O(w_2,\bar{w}_2)L_{-1}O(w_3,\bar{w}_3)L_{-1}O(w_4,\bar{w}_4)\rangle_{\Sigma_2}}{\left(\langle L_{-1}O(w_1,\bar{w}_1)L_{-1}O(w_2,\bar{w}_2)\rangle_{\Sigma_1}\right)^2} \nonumber\\
&& \simeq \frac{1}{(2\Delta)^2(2\Delta+1)^2{\mathcal{N}}^2}\cdot \frac{16\epsilon^{8\Delta+4}}{t^{4\Delta}} (\prod_{i=1}^4 N_i)\partial_{z_1}\partial_{z_2}\partial_{z_3}\partial_{z_4}\langle O(z_1,\bar{z}_1)O(z_2,\bar{z}_2)O(z_3,\bar{z}_3)O(z_4,\bar{z}_4)\rangle_{\Sigma_1} \nonumber\\
&& \simeq \frac{1}{(2\Delta^2)(2\Delta+1)^2{\mathcal{N}}^2}\cdot \frac{16\epsilon^{8\Delta+4}}{(4t^2)^{4\Delta}} \cdot (\prod_{i=1}^4 N_i)\partial_{z_1}\partial_{z_2}\partial_{z_3}\partial_{z_4} G(z,\bar{z}).
\label{ratio}
\eea

\subsubsection{Example I: $(2,1)$ operator in minimal model }

Now we would like to study an explicit examples: the $(2,1)$ operator in the minimal models.
Let us  consider a $(p,p')$ minimal model with $p>p'$, where the primary fields $\phi_{(r_1,r_2)}$ are specified by a pair of integers $(r_1,r_2)$ taking the values
\bea
r_1=1,2,...,p'-1, \ \ \  r_2=1,2,...,p-1.
\eea
The central charge of $(p,p')$ minimal model is
\bea
c=1-6\frac{(p-p')^2}{pp'},
\eea
and the conformal dimension of the $(r_1,r_2)$ primary operator is given by
\bea
\Delta_{r_1,r_2}=\frac{(pr_1-p'r_2)^2-(p-p')^2}{4pp'}.
\eea

Let us focus on the $(2,1)$ operator $\phi_{(2,1)}$ in a $(p,p')$ minimal model as it has a relatively simple four-point function. It has conformal dimension
\bea
\Delta_{(2,1)}=\frac{3p}{4p'}-\f{1}{2}.\ \
\eea
The function $G(z,\bar{z})$ for the four-point function $\langle \phi_{2,1}(z_1,\bar{z}_1)
... \phi_{2,1}(z_4,\bar{z}_4)\rangle_{\Sigma_1}$  via the relation (\ref{gf}) is known to be \cite{Dotsenko:1984nm}\cite{Dotsenko:1984ad}
\bea
&& G(z,\bar{z}) \nonumber\\
&& =|z|^{\frac{p}{p'}}|1-z|^{\frac{p}{p'}}\cdot \left[\f{\sin\left(\frac{\pi p}{p'}\right)\sin\left(\frac{3\pi p}{p'}\right)}{\sin\left(\frac{2\pi p}{p'}\right)}|I_1(z)|^2
+\frac{\sin\left(\frac{\pi p}{p'}\right)\sin\left(\frac{\pi p}{p'}\right)}{\sin\left(\frac{2\pi p}{p'}\right)}|I_{2}(z)|^2\right], \label{gfun}
\eea
where we have defined \cite{CFT}\cite{BPZ}
\bea
&& I_1(z)=\f{\Gamma\left(\f{3p}{p'}-1\right)\Gamma\left(1-\frac{p}{p'}\right)}
{\Gamma\left(\frac{2p}{p'}\right)}\cdot F\left[\frac{p}{p'},-1+\frac{3p}{p'},\frac{2p}{p'};z\right], \nonumber\\
&& I_2(z)=z^{1-\frac{2p}{p'}}\cdot\frac{\Gamma\left(1-\frac{p}{p'}\right)\Gamma\left(1-\frac{p}{p'}\right)}
{\Gamma\left(2-\frac{2p}{p'}\right)}\cdot F\left[\frac{p}{p'},1-\frac{p}{p'},2-\frac{2p}{p'};z\right].
\eea
with $F[a,b,c;z]$ being the hypergeometric function.
By taking the limit $z_{12}=z_{34}\to 0$, we find that
\bea
\langle \phi_{2,1}(z_1,\bar{z}_1)
... \phi_{2,1}(z_4,\bar{z}_4)\rangle_{\Sigma_1}
\to |z_{12}|^{-8\Delta}\cdot \frac{\sin\left(\frac{\pi p}{p'}\right)^2}{\sin\left(\frac{2\pi p}{p'}\right)}\cdot \frac{\Gamma\left(1-\f{p}{p'}\right)^4}
{\Gamma\left(2-\frac{2p}{p'}\right)^2}.
\eea
Thus we can identify the normalization factor ${{\cal N}}$ of the two-point function as follows
\bea
{{\cal N}}^2=\frac{\sin\left(\frac{\pi p}{p'}\right)^2}{\sin\left(\frac{2\pi p}{p'}\right)}\cdot \frac{\Gamma\left(1-\frac{p}{p'}\right)^4}
{\Gamma\left(2-\frac{2p}{p'}\right)^2}.
\eea


Now we turn to the most interesting limit: late time limit $t>>l>>\epsilon$, where we have
$z=1-\frac{\epsilon^2}{4t^2}$ and $\bar{z}=\frac{\epsilon^2}{4t^2}$. Under this limit we can put (\ref{gfun}) into the final formula (\ref{ratio}). Notice that  the $I_2$ term is  more divergent  in the small $\epsilon$ limit, and using the identities on the  hypergeometric functions, we have
\bea
&& I_2(z)\simeq (1-z)^{1-\frac{2p}{p^{'}}}\frac{\Gamma\left(1-\frac{p}{p'}\right)\Gamma\left(-1+\frac{2p}{p'}\right)}
{\Gamma\left(\frac{p}{p'}\right)}
=\frac{\Gamma\left(1-\frac{p}{p'}\right)\Gamma\left(-1+\frac{2p}{p'}\right)}
{\Gamma\left(\frac{p}{p'}\right)}\cdot \left(\frac{\epsilon^2}{4t^2}\right)^{1-2p/p'},\nonumber\\
&& I_2(\bar{z})\simeq
\bar{z}^{1-\frac{2p}{p^{'}}} \frac{\Gamma\left(1-\frac{p}{p'}\right)^2}
{\Gamma\left(2-\frac{p}{p'}\right)}
=\frac{\Gamma\left(1-\frac{p}{p'}\right)^2}
{\Gamma\left(2-\frac{p}{p'}\right)}\cdot \left(\frac{\epsilon^2}{4t^2}\right)^{1-2p/p'}.
\eea
Thus we find
\be
G(z,\bar{z})\simeq \f{\sin\left(\f{\pi p}{p'}\right)^2}{\sin\left(\f{2\pi p}{p'}\right)}
\cdot \f{\Gamma\left(1-\f{p}{p'}\right)^3\Gamma\left(-1+\f{2p}{p'}\right)}
{\Gamma\left(2-\f{2p}{p'}\right)\Gamma\left(\f{p}{p'}\right)}\cdot \left(\f{\ep^2}{4t^2}\right)^{2-3p/p'}.
\ee
In this way, we finally obtain
\be
\mbox{Tr}\rho_A^2=\f{\Gamma\left(2-\f{2p}{p'}\right)^3\Gamma\left(-1+\f{2p}{p'}\right)}
{\Gamma\left(1-\f{p}{p'}\right)\Gamma\left(\f{p}{p'}\right)}
=-\f{1}{2\cos\left(\f{\pi p}{p'}\right)},
\ee
and the Renyi entanglement entropy difference
\be
\Delta S^{(2)}_A=\log \left[-2\cos\left(\f{\pi p}{p'}\right)\right].
\ee
For a general minimal model, the entropy difference is not vanishing for the descendent operator $L_{-1}O(2,1)$ operator. But
for the Ising model $(p,p')=(4,3)$ \cite{CFT}\cite{BPZ}, it is vanishing.

In a minimal model, the modular transformation maps the
primary operator $(s_1,s_2)$ into $(r_1,r_2)$. Its S-matrix is defined to be $S_{(r_1,r_2),(s_1,s_2)}$, given explicitly by
\be
S_{(r_1,r_2),(s_1,s_2)}=2\s{\f{2}{pp'}}(-1)^{1+r_2s_1+r_1s_2}\sin\left(\f{\pi p}{p'}r_1s_1\right)\sin\left(\f{\pi p'}{p}r_2s_2\right).
\ee
The quantum dimension $d_{(r_1,r_2)}$ for the primary field $(r_1,r_2)$ is defined by
\be
d_{(r_1,r_2)}=\f{S_{(1,1),(r_1,r_2)}}{S_{(1,1),(1,1)}}=-2\cos\left(\f{\pi p}{p'}\right).
\ee
Thus we can conclude that for the $L_{-1}O(2,1)$ operator
\be
\Delta S^{(2)}_A=\log d_{(r_1,r_2)},
\ee
the same as the one for the primary operator $O(2,1)$.

\subsubsection{Example II: energy momentum tensor}

Another simple case is the excitation of energy momentum tenor, which is the descendent state of the identity operator. Different from the case discussed above, it is $L_{-2}$ rather than $L_{-1}$. We only consider $\Delta S_2$ here.
The two-, three- and four-point correlation functions on the $R^2$ depend only  on the central charge $c$ \cite{Osborn:2012vt}, which are respectively
\begin{eqnarray}
\langle T(z_1)T(z_2)\rangle =\frac{c}{2z_{12}^2},
\end{eqnarray}
\begin{eqnarray}\label{EMTensor4point}
\langle T(z_1)T(z_2)T(z_3)T(z_4) \rangle =\frac{\mathcal{F}(z)}{z_{12}^4z_{34}^4},
\end{eqnarray}
with
\begin{eqnarray}
\mathcal{F}(z)=\frac{c^4}{4}\Big(1+z^4+\frac{z^4}{(1-z)^4}\Big)+2c\frac{z^2(1-z+z^2)}{(1-z)^2},
\end{eqnarray}
$z=z_{12}z_{34}/(z_{13}z_{24})$ is the ratio. The transformation of the energy momentum tensor under the map $z(\omega)$ is given by
\begin{eqnarray}
T(\omega)=\Big(\frac{dz}{d\omega}\Big)^2T(z)+\frac{c}{12}\{z,\omega\},
\end{eqnarray}
with the Schwarzian derivative
\begin{eqnarray}
\{z,\omega\}=(z'''z'-\frac{3}{2}(z'')^2)/(z')^2.
\end{eqnarray}
In the coordinate (\ref{wcoordinate}) the leading order of the two-point function is
\begin{eqnarray}
\langle T(\omega_1) T(\omega_2) \rangle=\frac{1}{(2\epsilon)^4}.
\end{eqnarray}
After the transformation the four-point correlation function in the $\omega$ coordinate is
\begin{eqnarray}
\langle T(\omega_1)T(\omega_2)T(\omega_3)T(\omega_4) \rangle =\prod_i (\frac{dz_i}{d\omega_i})^2\langle T(z_1)T(z_2)T(z_3)T(z_4) \rangle+...,
\end{eqnarray}
where `...' denotes the less divergent terms, such as the product of three-point function and the Schwarzian derivative.
During the time $0<t<l$ or $t>L+l$, we can see $z_{12}\sim z_{34}\sim O(\epsilon)$, also $z\sim O(\epsilon^2)$. One could see from (\ref{EMTensor4point})
the leading contribution of the four-point function is
\begin{eqnarray}\label{EM4early}
\langle T(z_1)T(z_2)T(z_3)T(z_4) \rangle \simeq \frac{c^2}{4z_{12}^4z_{34}^4}.
\end{eqnarray}
with
\begin{eqnarray}
z_{12}=-\frac{iL\epsilon}{(t-L)(t-L-L)}z_1,\ \ z_{34}=-\frac{iL\epsilon}{(t-L(t-L-l))}.
\end{eqnarray}
During the time $l<t<L+l$, $z_1=-z_2$, $z_3=-z_4$, $z_{23}\sim z_{14}\sim O(\epsilon)$, and $z\sim 1-O(\epsilon^2)$, with
\begin{eqnarray}
z\simeq 1-\frac{L^2 \epsilon^2}{4(t-l)^2(t-L-l)^2}.
\end{eqnarray}
Now the leading order contribution of the four-point correlation function in the limit $\epsilon \to 0$ is
\begin{eqnarray}\label{EM4later}
\langle T(z_1)T(z_2)T(z_3)T(z_4) \rangle \simeq \frac{c^2}{4(1-z)^4} \frac{1}{(4z_{1}z_{3})^4}.
\end{eqnarray}
One could check (\ref{EM4early}) is the same as (\ref{EM4later}). Therefore, by using (\ref{replica}), we find $\Delta S^{(2)}=0$ during the evolution.
This is just as we expected, since $T(z)$ is the descendent operator of identity $I$, the quantum dimension of which is $1$.

\subsection{2nd R\'enyi entropy for some descendant operators}

In previous section, we computed the R\'enyi entropy of   some specific descendent operators and found that these descendant operators had the same contribution as the primary operators. In this subsection, we use the conformal block and operator product expansion (OPE) to show this is true for a large class of descendent operators.  It is also a warm up for the study of more general descendant operators in the next subsection. 

In this and next subsection, we will consider the case that the interval is finite, lying at $[0,L]$, and the operators is inserted at $-l$.
The two-point and $2n$-point functions on single sheet and $n$-sheeted Riemann surfaces are respectively
\be Z_1=\langle V(w,\bar{w})V(w^{'},\bar{w}^{'})\rangle,  \ee
\be Z_n=\langle V(w_1,\bar{w}_1) V(w^{'}_1,\bar{w}^{'}_1)  V(w_2,\bar{w}_2) V(w^{'}_2,\bar{w}^{'}_2)
... V(w_n,\bar{w}_n) V(w^{'}_n,\bar{w}^{'}_n)
 \rangle\mid_{n-sheets} ,\ee
where the correlation function $Z_n$ is defined on the $n$-sheeted surface, with $V(w_i,\bar{w}_i)$ on the $i$-th surface.
We set
\be w_i=-l+(t-i\epsilon), ~~~\bar{w}_i=-l+(t+i\epsilon), \ee
\be w_i^{'}=-l+(t+i\epsilon),~~~\bar{w}_i^{'}=-l-(t+i\epsilon). \ee
$V(w,\bar{w})$ can be any descendant operators.\footnote{For convenience the convention in this section and the next section is a little different from the one in other sections. The correspondence is $w_{2j}\rightarrow w_{j}$, $w_{2j-1}\rightarrow w'_j$} We only need to set $n=2$ for $S^{(2)}$ calculation. To evaluate the multi-point correlation function on $n$-sheeted surface, we need to take a conformal transformation
\be\label{ztow} z=(\frac{w}{w-L})^{\frac{1}{n}}. \ee

First we consider the descendent operator  $V(w,\bar{w})=\partial O_a(w,\bar{w})$, where $O_a(w,\bar{w})$ is a primary operator, the two-point function equals to
 \be Z_1=\langle \partial O_a(w,\bar{w}) \partial O_a(w^{'},\bar{w}^{'}) \rangle
 =-2h(2h+1)\frac{1}{(w-w^{'})^{2h+2}}\frac{1}{(\bar{w}-\bar{w}^{'})^{2h}}=\frac{2h(2h+1)}{(2\epsilon)^{4h+2}}, \ee
 and the four-point function is
 \bea Z_2
 =\partial_{w_1}\partial_{w_1^{'}}\partial_{w_2}\partial _{w_2^{'}}
 \langle  O_a(w_1,\bar{w}_1)O_a(w_1^{'},\bar{w}_1^{'})O_a(w_2,\bar{w}_2)O_a(w_2^{'},\bar{w}_2^{'})\rangle.
\eea
Under the conformal transformation (\ref{ztow}), for $t<l$
\be z_1=\frac{(l-t+i\epsilon)^{\frac{1}{2}}}{(l+L-t+i\epsilon)^{\frac{1}{2}}}, \hs{3ex}
 \bar{z}_1=\frac{(l+t-i\epsilon)^{\frac{1}{2}}}{(L+l+t-i\epsilon)^{\frac{1}{2}}}, \ee
\be z_1^{'}=\frac{(l-t-i\epsilon)^{\frac{1}{2}}}{(l+L-t-i\epsilon)^{\frac{1}{2}}}, \hs{3ex}
 \bar{z}_1^{'}=\frac{(l+t+i\epsilon)^{\frac{1}{2}}}{(l+L+t+i\epsilon)^{\frac{1}{2}}} \ee
\be z_2=-\frac{(l-t+i\epsilon)^{\frac{1}{2}}}{(l+L-t+i\epsilon)^{\frac{1}{2}}} \hs{3ex}
 \bar{z}_2=-\frac{(l+t-i\epsilon)^{\frac{1}{2}}}{(L+l+t-i\epsilon)^{\frac{1}{2}}} \ee
\be z_2^{'}=-\frac{(l-t-i\epsilon)^{\frac{1}{2}}}{(l+L-t-i\epsilon)^{\frac{1}{2}}} \hs{3ex}
 \bar{z}_2^{'}=-\frac{(l+t+i\epsilon)^{\frac{1}{2}}}{(l+L+t+i\epsilon)^{\frac{1}{2}}}, \ee
where $(z_1,z_1^{'})$ $(z_2,z_2^{'})$ $(\bar{z}_1,\bar{z}_1^{'})$ $(\bar{z}_2,\bar{z}_2^{'})$ are close to each other. While for $l<t<l+L$,
\be z_1=-z_2=e^{\frac{\pi i}{2}}\frac{(t-l-i\epsilon)^{\frac{1}{2}}}{(l+L-t+i\epsilon)^{\frac{1}{2}}}, \ee
\be z_1^{'}=-z_2^{'}=e^{-\frac{\pi i}{2}}\frac{(t-l+i\epsilon)^{\frac{1}{2}}}{(l+L-t-i\epsilon)^{\frac{1}{2}}}, \ee
then  $(z_1,z_2^{'})$ $(z_2,z_1^{'})$ $(\bar{z}_1,\bar{z}_1^{'})$ $(\bar{z}_2,\bar{z}_2^{'})$ are close to each other.

From the result in \cite{He:2014mwa},  for $t<l$,
\bea\lefteqn{\langle O_a(z_1,\bar{z}_1)O_a(z_1^{'},\bar{z}_1^{'})O_a(z_2,\bar{z}_2)O_a(z_2^{'},\bar{z}_2^{'}) \rangle} \notag \\
&=&\frac{1}{(z_1-z_1^{'})^{2h}(\bar{z}_1-\bar{z}_1^{'})^{2h}}\frac{1}{(z_2-z_2^{'})^{2h}(\bar{z}_2-\bar{z}_2^{'})^{2h}}
+\mbox{less divergent term},
\eea
and for $l<t<l+L$,
\bea \lefteqn{\langle O_a(z_1,\bar{z}_1)O_a(z_1^{'},\bar{z}_1^{'})O_a(z_2,\bar{z}_2)O_a(z_2^{'},\bar{z}_2^{'}) \rangle} \notag \\
&=&\frac{1}{d_a}
\frac{1}{(z_1-z_2^{'})^{2h}(\bar{z}_1-\bar{z}_1^{'})^{2h}}\frac{1}{(z_2-z_1^{'})^{2h}(\bar{z}_2-\bar{z}_2^{'})^{2h}}
+\mbox{less divergent term}.
\eea
Taking a conformal transformation (\ref{ztow}), for $t<l$
\bea \lefteqn{\langle O_a(w_1,\bar{w}_1)O_a(w_1^{'},\bar{w}_1^{'})O_a(w_2,\bar{w}_2)O_a(w_2^{'},\bar{w}_2^{'}) \rangle
\mid_{2-sheets}}
 \notag \\
&=&\frac{1}{(w_1-w_1^{'})^{2h}(\bar{w}_1-\bar{w}_1^{'})^{2h}}\frac{1}{(w_2-w_2^{'})^{2h}(\bar{w}_2-\bar{w}_2^{'})^{2h}}
+\mbox{less divergent term},
\eea
and for $l<t<l+L$
\bea
\lefteqn{\langle O_a(w_1,\bar{w}_1)O_a(w_1^{'},\bar{w}_1^{'})O_a(w_2,\bar{w}_2)O_a(w_2^{'},\bar{w}_2^{'}) \rangle
\mid_{2-sheets}}
 \notag \\
&=& \frac{1}{d_a}
\frac{1}{(w_1-w_2^{'})^{2h}(\bar{w}_1-\bar{w}_1^{'})^{2h}}\frac{1}{(w_2-w_1^{'})^{2h}(\bar{w}_2-\bar{w}_2^{'})^{2h}}
+\mbox{less divergent term},
\eea
where we still only keep the leading divergent term for small $\epsilon$.
Consequently for $l<t<l+L$
\bea
\lefteqn{\partial_{w_1}\partial_{w_1^{'}}\partial_{w_2}\partial_{w_2^{'}}
\langle O_a(w_1,\bar{w}_1)O_a(w_1^{'},\bar{w}_1^{'})O_a(w_2,\bar{w}_2)O_a(w_2^{'},\bar{w}_2^{'}) \rangle }\notag \\
&=&\frac{(2h)^2(2h+1)^2}{d_a}
\frac{1}{(w_1-w_2^{'})^{2h+2}(\bar{w}_1-\bar{w}_1^{'})^{2h}}\frac{1}{(w_2-w_1^{'})^{2h+2}(\bar{w}_2-\bar{w}_2^{'})^{2h}}
+\mbox{lower divergent term} \notag \\
&=&\frac{(2h)^2(2h+1)^2}{d_a}\frac{1}{(2\epsilon)^{8h+4}}+\mbox{lower divergent term},
\eea
which leads to $\Delta S_2=\log d_a$ for $l<t<l+L$.

Next we consider  the descendent operators with the form
\be V(w,\bar{w})=L^{(-)}O(w,\bar{w}), \ee
where $L^{(-)}$ is a  combination of holomorphic generators such that $V(w,\bar{w})$ is a quasi-primary operator\footnote{This condition can be relaxed from the result in the next subsection, but for simplicity we still assume this condition in this subsection.}. We also assume the operator $L^{(-)}$ has a fixed conformal dimension
\be [L_0,L^{(-)}]=mL^{(-)}. \ee
Because the final result only depend on the most singular term in the two-point function and the $2n$-point function, there is nothing change by adding in the  operators with smaller conformal dimensions.

The conformal transformation for the descendant operators are different from the one for the primary operator.  Under a conformal transformation
\bea L_{-m}\mid_{w_1}&=&\frac{1}{2\pi i} \oint \frac{dw}{(w-w_1)^{m-1}}T(w) \notag \\
&=&\frac{1}{2\pi i}\oint \frac{dw}{(w-w_1)^{m-1}}(T(z)(\frac{\partial z}{\partial w})^2
+\frac{c}{12}\{z,w\}) \notag \\
&=&\frac{1}{2\pi i}\oint \frac{dw}{(w-w_1)^{m-1}}
(\sum_{r}\frac{L_{-r}\mid_{z_1}}{(z-z_1)^{-r+2}}
(\frac{\partial z}{\partial w})^2
+\frac{c}{12}\{z,w\}) \notag \\
&=&L_{-m}\mid_{z_1}(\frac{\partial z_1}{\partial w_1})^n+...
\eea
Here the ellipsis denote the terms with lower conformal dimensions, which leads to less  divergent terms in the limit  $\epsilon\rightarrow0$.\footnote{The most divergent term in the limit $\epsilon\rightarrow 0$ comes from the OPE of two operators to the identity operator. }
Under the conformal transformation (\ref{ztow}), the four-point function transforms as
\bea\label{tra}
 \lefteqn{\langle V(w_1,\bar{w}_1)V(w^{'}_1,\bar{w}_1^{'})V(w_2,\bar{w}_2)V(w_2^{'},\bar{w}_2^{'})\rangle} \notag \\
&=&(\frac{\partial w_1}{\partial z_1})^{h+m}(\frac{\partial \bar{w}_1}{\partial \bar{z}_1})^{h+\bar{m}}
(\frac{\partial w_2}{\partial z_2})^{h+m}(\frac{\partial \bar{w}_2}{\partial \bar{z}_2})^{h+\bar{m}}
(\frac{\partial w_1^{'}}{\partial z_1^{'}})^{h+m}(\frac{\partial \bar{w}_1^{'}}{\partial \bar{z}_1^{'}})^{h+\bar{m}}
(\frac{\partial w_2^{'}}{\partial z_2^{'}})^{h+m}(\frac{\partial \bar{w}_2^{'}}{\partial \bar{z}_2^{'}})^{h+\bar{m}}
\notag \\
&~&\cdot \langle V(z_1,\bar{z}_1)V(z^{'}_1,\bar{z}_1^{'})V(z_2,\bar{z}_2)V(z_2^{'},\bar{z}_2^{'})\rangle
+\mbox{less divergent terms}.
\eea
Even though the operator $V(w,\bar{w})$ is not a primary operator, the coefficient for the leading  term is the same as the one for the primary operator.
The terms with lower conformal dimensions are less divergent  in the small $\epsilon$ limit, so do not contribute to the final result.

Because we assume that the operator $V$ is a quasi-primary operator, it transforms homogeneously under a linear conformal transformation. Consider the conformal transformation
\be u=\frac{1}{z} \ee
we have
\bea \lefteqn{\lim_{z\rightarrow \inf} z^{2(h+m)}L^{(-)}O(z) }\notag \\
&=&\lim_{{z\rightarrow \inf}}(\frac{\partial u}{\partial z})^{(h+m)} z^{2(h+m)} L^{(-)}O(u) \notag \\
&=&(-1)^{h+m} L^{(-)}O(u)\mid_{u=0}.
\eea
Furthermore
\bea L_{-n}\mid_{u=0}&=&\frac{1}{2\pi i}\oint \frac{du}{u^{n-1}}T(u) \notag \\
&=&\frac{1}{2\pi i}\oint \frac{dz}{z^{-n+3}}T(z)z^{4} \notag \\
&=&L_n\mid_{z\rightarrow \inf}
\eea
and
\be  \lim_{z\rightarrow \inf} z^{2(h+m)}\langle 0\mid L^{(-)}O(z)=\langle O\mid L^{(-)\dag}(-1)^{m}, \ee
then the  two-point function is
\be\label{twopoint} \langle L^{(-)}O(z),L^{(-)}O(z^{'}) \rangle =\frac{c_0}{(z-z^{'})^{2(h+m)}}, \ee
where $c_0=(-1)^m\langle h \mid L^{(-)\dag}L^{(-)} \mid h \rangle $.

Now let us  study the divergence in the correlation function of four descendants in the $z$ coordinate. We set $l<t<l+L$ then $(z_1,z_2^{'})$ $(z_2,z_1^{'})$ $(\bar{z}_1,\bar{z}_1^{'})$ $(\bar{z}_2,\bar{z}_2^{'})$ are close to each other.
The four-point correlation function of quasi-primary operators can be transformed into
\bea \lefteqn{\langle V(z_1,\bar{z}_1)V(z_1^{'},\bar{z}_1^{'})V(z_2,\bar{z}_2)V(z_2^{'},\bar{z}_2^{'}) \rangle }\notag \\
&=&{\cal{D}}
\langle O(z_1,\bar{z}_1)O(z_1^{'},\bar{z}_1^{'})O(z_2,\bar{z}_2)O(z_2^{'},\bar{z}_2^{'}) \rangle \notag \\
&=&{\cal{D}}
\sum_m c_{m} \langle O(z_1)O(z_1^{'})\mid_m O(z_2)O(z_2^{'}) \rangle
\langle O(\bar{z}_1)O(\bar{z}_1^{'})\mid_m O(\bar{z}_2) O(\bar{z}_2^{'}) \rangle \notag \\
&=& {\cal{D}}
\sum_{m,n} c_{m,n} \langle O(z_1)O(z_2^{'})\mid_m O(z_2)O(z_1^{'}) \rangle
\langle O(\bar{z}_1)O(\bar{z}_1^{'})\mid_n O(\bar{z}_2) O(\bar{z}_2^{'}) \rangle \notag \\
&=&\sum_{m,n} c_{m,n} \langle L^{(-)}O(z_1)L^{(-)}O(z_2^{'})\mid_m L^{(-)}O(z_2)L^{(-)}O(z_1^{'}) \rangle
\langle O(\bar{z}_1)O(\bar{z}_1^{'})\mid_n O(\bar{z}_2) O(\bar{z}_2^{'}) \rangle, \notag \\
\eea
where $\langle O(z_1)O(z_1^{'})\mid_mO(z_2)O(z_2^{'})\rangle$ denote the conformal block expansion with the  Virasoro module $[m]$ as the propagator.
The first equation transforms the correlation function of four descendants  into the differential on the correlation function of corresponding primaries. In the second equation, we expand the partition function by the conformal blocks and $c_m$ denote the OPE coefficient. In the third equation, we expand the holomorphic part in terms of another channel \cite{He:2014mwa}. In the fourth equation, we change the differential back into the Virasoro operators acting on the primaries  in the correlation function. The last equation is based on the fact that the Ward identity is satisfied for the conformal blocks. 

The most divergent term only comes from the one with $m=n=0$ and $c_{0,0}=\frac{1}{d_a}$ \cite{He:2014mwa}. Actually even in the vacuum block, only the identity operator gives the most divergent term,
\bea \lefteqn{\langle V(z_1,\bar{z}_1)V(z_1^{'},\bar{z}_1^{'})V(z_2,\bar{z}_2)V(z_2^{'},\bar{z}_2^{'}) \rangle} \notag \\
&=& \frac{1}{d_a}
\langle L^{(-)}O(z_1)L^{(-)}O(z_2^{'})\rangle \langle L^{(-)}O(z_2)L^{(-)}O(z_1^{'}) \rangle
\langle O(\bar{z}_1)O(\bar{z}_1^{'}) \rangle \langle O(\bar{z}_2) O(\bar{z}_2^{'}) \rangle\nonumber\\&~&
+\mbox{less divergent terms.}
\eea
Changing back into the $w$-coordinate and keeping the most divergent term, we find
\bea \lefteqn{\langle V(w_1,\bar{w}_1)V(w_1^{'},\bar{w}_1^{'})V(w_2,\bar{w}_2)V(w_2^{'},\bar{w}_2^{'}) \rangle }\notag \\
&=& \frac{1}{d_a}
\langle L^{(-)}O(w_1)L^{(-)}O(w_2^{'})\rangle \langle L^{(-)}O(w_2)L^{(-)}O(w_1^{'}) \rangle
\langle O(\bar{w}_1)O(\bar{w}_1^{'}) \rangle \langle O(\bar{w}_2) O(\bar{w}_2^{'})  \rangle \notag \\
&~&+\mbox{less divergent terms.}
\eea
Therefore, for a quasi-primary operator we still have $\Delta S_2=\log d_a$.

In short, for two kinds of descendent operators: $\p O$ and $L^{(-)}O$, the difference between the second R\'enyi entropy of the excited states and the ground state is universal, equals to the logarithmic of the quantum dimension. This is the same as the case of inserting a local primary operator.

\subsection{$n$-th RE for generic descendent states}

With the above study, we are ready to discuss the effect of most general descendent operators. A descendent operator in a module generated from a primary operator $O_a$ may take the following generic form
\be V=\sum_{m,j,r,k}d_{m,j;r,k}(\partial^{m}L^{(-,j)})(\bar{\partial}^{r}\bar{L}^{(-,k)} )O_a(w,\bar{w}) \ee
where $L^{(-,j)}\bar{L}^{(-,k)}O_a(w,\bar{w})$ is a quasi-primary operator, and $L^{(-,j)}$ ($\bar{L}^{(-,k)}$) is a combination of holomorphic (anti-holomorphic) Virasoro algebra with fixed conformal dimension $[L_0,L^{(-,j)}]=p_jL^{(-,j)}$.
Keeping the most divergent term in the correlation function, we see that only the terms with $m+r+p_j+\bar{p}_k+2h_a=\Delta$ is non-zero. For general $n$-sheeted surface under the conformal transformation (\ref{ztow})
\be z_j=\frac{(t-l-i\epsilon)^{\frac{1}{n}}}{(l+L-t+i\epsilon)^{\frac{1}{n}}}
e^{\frac{2\pi i}{n}(j-1)+\frac{\pi i}{n}} ,\ee
\be z_j^{'}=\frac{(t-l+i\epsilon)^{\frac{1}{n}}}{(l+L-t-i\epsilon)^{\frac{1}{2}}}
e^{\frac{2\pi i}{n}(j-1)-\frac{\pi i}{n}} ,\ee
we see that $z_j$ is close to $z_{j+1}^{'}$ as in \cite{He:2014mwa}.

\begin{figure}
  \centering
  \includegraphics[width=8cm]{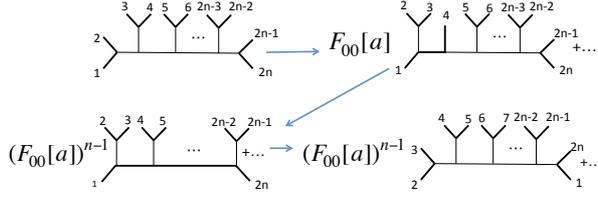}\\
  \caption{The fusion transformations to obtain $\Delta S^{(n)}_A$.}\label{conformalblock}
\end{figure}

For the $n$-th R\'enyi entropy, we need to compute the $2n$-point function. Similar to the previous calculation we need to take a conformal transformation to $z$ coordinate and take proper channel to expand the $2n$-point function into the holomorphic and the anti-holomorphic part, as graphically shown in Fig. \ref{conformalblock}\footnote{We have extracted this figure from\cite{He:2014mwa}.}. In each channel only the identity operator contribute to the final result, so the $2n$-point function breaks up into $n$ two-point functions for the holomorphic part (and $n$ for the anti-holomorphic part). We can take a conformal transformation back to the $w$ coordinate, with the leading divergent term being transformed homogenously. Actually this recipe only needs us to {know which} two arguments are close to each other and the coefficients for the most divergent term. With this message we can directly do the calculation in the $w$ coordinate and multiply the coefficients from the OPE and the channel changing
\bea \lefteqn{\langle V(w_1,\bar{w}_1)V(w_1^{'},\bar{w}_1^{'})...V(w_n,\bar{w}_n)V(w^{'}_n,\bar{w}^{'}_n) \rangle
\mid_{n-sheet}}\notag \\
&=&\sum_{m_1,j_1,r_1,k_1}\sum_{m_1^{'},j_1^{'},r_1^{'},k_1^{'}}
d_{m_1,j_1;r_1,k_1}d^{*}_{m_1^{'},j_1^{'};r_1^{'},k_1^{'}}...
\sum_{m_n,j_n,r_n,k_n}\sum_{m_n^{'},j_n^{'},r_n^{'},k_n^{'}}
d_{m_n,j_n;r_n,k_n}d^{*}_{m_n^{'},j_n^{'};r_n^{'},k_n^{'}} \notag \\
&~&\langle \partial_{w_1}^{m_1}L^{(-,j_1)}\bar{\partial}_{\bar{w}_1}^{r_1}\bar{L}^{(-,k_1)}
O_a(w_1,\bar{w}_1)
\partial_{w_1^{'}}^{m_1^{'}}L^{(-,j_1^{'})}\bar{\partial}_{\bar{w_1}^{'}}^{r_1^{'}}\bar{L}^{(-,k_1^{'})}
O_a(w_1^{'},\bar{w}_1^{'})... \notag \\
&~& \partial_{w_n}^{m_n}L^{(-,j_n)}\bar{\partial}_{\bar{w}_n}^{r_n}\bar{L}^{(-,k_n)}
O_a(w_n,\bar{w}_n)
\partial_{w_n^{'}}^{m_n^{'}}L^{(-,j_n^{'})}\bar{\partial}_{\bar{w_n}^{'}}^{r_n^{'}}\bar{L}^{(-,k_n^{'})}
O_a(w_n^{'},\bar{w}_n^{'})\rangle \mid_{n-sheet}
 \notag \\
&=&d_a^{-(n-1)}\sum_{m_1,j_1,r_1,k_1}\sum_{m_1^{'},j_1^{'},r_1^{'},k_1^{'}}
d_{m_1,j_1;r_1,k_1}d^{*}_{m_1^{'},j_1^{'};r_1^{'},k_1^{'}}...
\sum_{m_n,j_n,r_n,k_n}\sum_{m_n^{'},j_n^{'},r_n^{'},k_n^{'}}
d_{m_n,j_n,r_n,k_n}d^{*}_{m_n^{'},j_n^{'},r_n^{'},k_n^{'}} \notag \\
&~&\langle \partial_{w_1}^{m_1}L^{(-,j_1)}O_a(w_1)\partial_{w_2^{'}}^{m_2^{'}}L^{(-,j_2^{'})}O_a(w_2^{'})\rangle\cdots \langle \partial_{w_n}^{m_n}L^{(-,j_n)}O_a(w_n)\partial_{w_1^{'}}^{m_1^{'}}L^{(-,j_1^{'})}O_a(w_1^{'})\rangle\notag\\
&~&\langle \partial_{\bar{w_1}}^{r_1}\bar{L}^{(-,k_1)}O_a(\bar{w_1})
\partial_{\bar{w}_1^{'}}^{r_1^{'}}\bar{L}^{(-,k_1^{'})}O_a(\bar{w}_1^{'}) \rangle ...
\langle \partial_{\bar{w_n}}^{r_n}\bar{L}^{(-,k_n)}O_a(\bar{w_n})
\partial_{\bar{w}_n^{'}}^{r_n^{'}}\bar{L}^{(-,k_n^{'})}O_a(\bar{w}_n^{'}) \rangle \notag \\
&~&+\mbox{less divergent terms}. \label{2npoint}
\eea
Each two-point function in the above relation can be computed directly. For example, for the holomorphic part, it is
\bea
\lefteqn{ \langle \partial_{w_i}^{m_i}L^{(-,j_i)}O_a(w_i)
\partial_{w_{i+1}^{'}}^{m_{i+1}^{'}}L^{(-,j_{i+1}^{'})}O_a(w_{i+1}^{'})\rangle} \notag \\
&=&
\partial_{w_i}^{m_i}\partial_{w_{i+1}^{'}}^{m_{i+1}^{'}}
\frac{\langle h \mid L^{(-,j_i)\dag}L^{(-,j_i)}\mid h\rangle \delta_{j_i,j_{i+1}^{'}}(-1)^{p_{j_i}}}
{(w_i-w_{i+1}^{'})^{2(h+p_{j_i})}} \notag \\
&=& \frac{\langle h \mid L^{(-,j_i)\dag}L^{(-,j_i)}\mid h\rangle \delta_{j_i,j_{i+1}^{'}}(-1)^{p_{j_i}+m_{i}}}
{(w_i-w_{i+1}^{'})^{2(h+p_{j_i})+m_i+m_{i+1}^{'}}},\label{2.77}
\eea
where we have used (\ref{twopoint}) in the first equation.
Introducing the matrices
\be B_{\{m,j\},\{r,k\}}=d^{*}_{m,j,r,k}, \ee
\be M_{\{m,j\},\{r,k\}}=\langle h\mid L^{(-,j)\dag}L^{(-,j)} \mid h\rangle \delta_{j,k}i^{r-m}, \ee
and defining the density matrix
\be \rho=BMB^{\dag}M^{\dag}, \ee
then
the two-point and $2n$-point function can be written as
\be \langle V(w_1,\bar{w}_1)V(w_1^{'},\bar{w}_1^{'})\rangle=\frac{tr \rho}{(2\epsilon)^{2\Delta}}, \ee
\be \langle V(w_1,\bar{w}_1)V(w_1^{'},\bar{w}_1^{'})...V(w_n,\bar{w}_n)V(w^{'}_n,\bar{w}^{'}_n) \rangle
\mid_{n-sheet}\notag \\
=\frac{tr \rho^n}{(2\epsilon)^{2n\Delta}}. \ee
With the normalized density matrix
\be \rho_0=\frac{\rho}{tr \rho},  \ee
we find
\be \frac{Z_n}{Z_1^n}=d_a^{-(n-1)}tr \rho_0^{n},  \ee
and therefore
\be\label{DecendentResult1} \Delta S_{n}=\log d_a-\frac{1}{n-1}\log tr \rho_0^{n}=\Delta S_n^{\mbox{\tiny primary}}-\frac{1}{n-1}\log tr \rho_0^{n}, \ee
and
\be\label{DecendentResult} \Delta S_{EE}=\log d_a-tr \rho_0 \log \rho_0=\Delta S_{EE}^{\mbox{\tiny primary}}
-tr \rho_0 \log \rho_0\ee
where
\be
\Delta S_n^{\mbox{\tiny primary}}=\Delta S_{EE}^{\mbox{\tiny primary}}=\log d_a
\ee
is the quantum entanglement of the primary operator.

It is obvious that there are two kinds of contribution to the entropy of a generic descendent operator. The first one takes a universal form, depending on the quantum dimension of the corresponding primary operator. Even though the theory could be different such that the OPE coefficients and the channel changing coefficients are different,   the $2n$-point function always takes the form  (\ref{2npoint}) with a different coefficient, so the relation between the entropies of the descendant operator and the corresponding primary operator does not change.

The extra contribution to the entropy is remarkable.  Formally it takes the form $-tr\rho_0\log \rho_0$. It is zero only when the matrix $d_{m,j;r,k}$ is of rank one. By matrix product, the matrix $\rho_0$ is also of rank one. under a conjugate transformation, $\rho_0$ can be diagnosed as $(1,0,0...)$, which has no more correction to the  entanglement entropy and R\'enyi entropy. In this case the entanglement and R\'enyi entropy are equal to the log of quantum dimension. The two examples in the previous section $\partial O(w,\bar{w})$ and $L^{(-)}O(w,\bar{w})$ belong to this class. A general form of this kind operator is like
\be\label{holomorphic} V(w,\bar{w})={\cal{L}}^{(-)}\bar{\cal{L}}^{(-)}O(w,\bar{w}), \ee
where $\cal{L}$ and $\bar{\cal{L}}$ are composed of holomorphic and anti-holomorphic Virasoro generators respectively.

For the cases when the rank of matrix $d_{m,j;r,k}$ is more than one, the entanglement entropy and R\'enyi entropy have extra corrections. The simplest example is for $d_{1,0;0,0}=d_{0,0;1,0}=1$ with other coefficients being zero. This gives the operator $V=(\partial+\bar{\partial})O(w,\bar{w})$. The entanglement entropy has extra $\log 2$ correction.

The extra increase for the entanglement entropy is easy to understand in a free theory. For the free theory, the entanglement entropy for local operator can be understood by the quasi-particle. The increase of the entanglement entropy of the local operator from the vacuum is equal to the entanglement of the EPR pair \cite{Nozaki:2014hna}. The entanglement entropy for the EPR pair only depends on the number of the pair and their relative normalizations. For the operator (\ref{holomorphic}), we change all of the left- and right-moving particles simultaneously. Because all of the EPR pairs have the same conformal dimension their normalizations change in the same way, which do not change the entanglement entropy. However for a more generic descendant operators which can not  decomposed as (\ref{holomorphic}), it  change the relative normalization of the EPR pair and even increase the number of EPR pairs. This is the origin of the extra increase in the entropies for the descendants. 

 From the derivation, we see that only the leading divergent term in the OPE  appears in the final result. The only thing we should consider is which operators are close to each other under the analytically extension
\be w \rightarrow -l+(t-i\epsilon) ,\text{ }\text{ }~\bar{w}\rightarrow -l-(t-i\epsilon). \ee
The crucial point here is that the holomorphic and anti-holomorphic parts have different limits. For example for the four-point function $\langle O_1(z_1,\bar{z}_1)O_2(z_2,\bar{z}_2)O_3(z_3,\bar{z}_3)O_4(z_4,\bar{z}_4)\rangle $, the $\epsilon\rightarrow 0$ limit set $(\bar{z}_1,\bar{z}_2)$ close to each other while $(z_1,z_3)$ close to each other. The OPE cannot be used directly. That's the reason why we need to transform into the $z$ coordinate and use the relation between the conformal blocks in different channel.


\subsection{Comments in BCFT}
In this subsection, we would like to introduce a boundary (or defect) at $x=0$ as shown in Fig. 1. This boundary preserves the conformal symmetry. There are two kinds of boundary preserving conformal symmetry, and they correspond to the Neumann boundary condition and Dirichlet boundary condition normally. The global property of the CFT with a boundary has been discussed in \cite{Calabrese:2005in}\cite{Cardy1}\cite{Cardy2}. As we know, the correlation function in a BCFT are much different from the one in CFT without boundary. In terms of (\ref{replica}), the R\'enyi entropy may be sensitive to the boundary. In \cite{Guo:2015uwa},  the R\'enyi entropy of the primary states with a boundary  has been  studied and it was shown that the maximal value of R\'enyi entropy does not change. The boundary effects just change the time evolution of R\'enyi entropy. It is also interesting to check what will happen to the R\'enyi entropy for the local descendent states in 2D CFT with a boundary. For simplicity, we only discuss the descendent operator whose entropy is the same as the one of corresponding primary operator.

As shown in \cite{He:2014mwa}\cite{Guo:2015uwa}, the R\'enyi entropy highly depends on the conformal blocks of the theory for general rational CFTs in 2D. In terms of \cite{Cardy1}\cite{Cardy2}, the $n$-point correlation functions in 2D CFTs with a boundary are related to the holomorphic part of the conformal blocks of the $2n$-point correlation functions on the 2D full complex plane. There is a systematical way called the image method to translate $n$-point correlation functions in 2D BCFTs to $2n$-point correlation functions in 2D CFT. The more precise relation is that an $n$-point function in the upper half plane(UHP), which is a function of the coordinates $(z_1,,z_n; \bar{z}_1,...,\bar{z}_n)$,
behaves under conformal transformations in the similar way as the holomorphic sector of an $2n$-point function in the full plane
which depends on $(z_1,...,z_n; z_1^*,...,z_n^*)$, analytically continued to $z_j^*=\bar{z_j}$. In \cite{He:2014mwa}, the time evolution of R\'enyi entropy highly depends on the holomorphic part of conformal block. In 2D CFTs with a boundary, the boundary indeed changes the propagator of the primary field  but does not change the fusion constants in the bulk. In this sense, one can expect that the time evolution of R\'enyi entropy of the descendent states in CFTs with a boundary is almost the same as the one in CFTs on the full complex plane. From the studies in the previous subsections, the R\'enyi entropy of the descendent states could be the same as the one of corresponding primary states. One can check that the boundary  does not change the maximal value of the R\'enyi entropy of the descendent states. In terms of the quasi-particles picture given in \cite{Guo:2015uwa}, the boundary just changes the time evolution of R\'enyi entropy of the descendent states, just  as  the primary states.

\section{ R\'enyi Entropy in Deformed CFT }

In the 2D rational CFT, we find that the  R\'enyi entropy of local descendent operators could  coincide with the logarithmic of the quantum dimension of the corresponding primary operator.
Now we would like to study these operators in the CFTs with additional deformations or interactions.
In the deformed CFTs, there is no  conformal invariance. Nevertheless the
effect of deformations on the R\'enyi entropy of local excited states can then be studied within conformal
perturbation theory \cite{Cardy:2010zs}, if the deformation is weak. Here we only  consider  the theory which is perturbed by the
local interaction $\Phi(z)$. Namely, we
 consider the theory which is perturbed by an operator $\Phi(z)$ with the conformal dimension $\Delta \ge 2$.  The action is now
\begin{eqnarray}\label{Actionperturbation}
I=I_0+\delta I= I_{0}+\lambda \int_{R_1} \Phi(z) d^2z,
\end{eqnarray}
where $I_0$ is the original CFT action, $\lambda \equiv g/\Lambda^{2-\Delta}$ with $g$ being dimensionless and $R_1$ denotes the complex plane. Consequently the total  Hamiltonian $H$ of the theory is  $H=H_{0}+\delta H$, where  $H_{0}$ is the Hamiltonian of  original  CFT. The system will be evolving under the Hamiltonian $H$ in the real time approach. As usual, we consider
an excited state by acting a primary or descendent operator $O$ on the vacuum $\ket{0}$, which is defined\footnote{We set the vacuum energy to be zero for convenience.} by $H\ket{0}=0$.

We consider the parameter $g \ll 1$ in (\ref{Actionperturbation}).
The two-point correlation function of the primary or the descendent operator is
\begin{eqnarray}\label{2pointPerturbation}
\langle O^{\dag}(\omega_1,\bar \omega_1)O(\omega_2,\bar \omega_2)\rangle_{\Sigma_1}= \langle e^{-\lambda \int_{R_1} \Phi(z) d^2z} O^{\dag}(\omega_1,\bar \omega_1)O(\omega_2,\bar \omega_2)\rangle_{R_1},
\end{eqnarray}
where ${\Sigma_1}$ and $R_1$ denote  the complex plane in deformed CFT and original CFT respectively. We expand the deformation with respect to the powers of $\lambda$
\begin{eqnarray}
\langle e^{-\lambda \int \Phi(z) d^2z}...\rangle_{R_1}=\sum_{N=0}^{+\infty} \frac{(-\lambda)^N}{N!}\int_{R_1}...\int_{R_1} \langle \Phi(z_1)... \Phi(z_N)...\rangle_{R_1}d^2z_1...d^2z_N.
\end{eqnarray}
The two-point correlation function (\ref{2pointPerturbation}) contains infinite towers of contribution from interaction between the operators $O$ and
 $\Phi$ in CFT. Similarly, the 2n-point correlation function on $\Sigma_n$ can be expanded
 \begin{eqnarray}
 &&\langle O^{\dag}(\omega_1,\bar \omega_1)O(\omega_2,\bar \omega_2)...O(\omega_{2n},\bar \omega_{2n})\rangle_{\Sigma_n}\nonumber \\
 &&=\langle e^{-\lambda \int \Phi(z) d^2z}O^{\dag}(\omega_1,\bar \omega_1)O(\omega_2,\bar \omega_2)...O(\omega_{2n},\bar \omega_{2n})\rangle_{R_n}\nonumber \\
 &&=\sum_{N=0}^{+\infty} \frac{(-\lambda)^N}{N!}\int_{R_n}...\int_{R_n} \langle \Phi(z_1)... \Phi(z_N)O^{\dag}(\omega_1,\bar \omega_1)O(\omega_2,\bar \omega_2)...O(\omega_{2n},\bar \omega_{2n})\rangle_{R_n}d^2z_1...d^2z_N,\nonumber
 \end{eqnarray}
 where $\Sigma_n$ is n copies of $\Sigma_1$.
 To calculate the correlation function on $R_n$, one can employ the following  conformal transformation
 \begin{eqnarray}\label{conformal transformation}
 \xi=\left(\frac{\omega}{\omega-l}\right)^{1/n}\text{and}\ \ \ \omega=\omega(\xi)=l\frac{\xi^n}{\xi^n-1}.
 \end{eqnarray}
 Then $R_n$ is mapped to the complex plane $R_1$. The points $\omega_1,...,\omega_{2n}$ is mapped to $z_1,...,z_{2n}$ respectively by (\ref{conformal transformation}). The $2n$-point correlation function can be expressed as
 \begin{eqnarray}
 &&\langle O^{\dag}(\omega_1,\bar \omega_1)O(\omega_2,\bar \omega_2)...O(\omega_{2n},\bar \omega_{2n})\rangle_{\Sigma_n}\nonumber \\
 &&=\sum_{N=0}^{+\infty}C_n \frac{(-\lambda)^N}{N!}\int_{R_1}...\int_{R_1} (\omega'(\tilde \xi_1)...\omega'(\tilde \xi_N))^{2-\Delta} \langle \Phi(\tilde \xi_1)... \Phi(\tilde \xi_N)\nonumber \\
 &&O^{\dag}(\xi_1,\bar \xi_1)O(\xi_2,\bar \xi_2)...O(\xi_{2n},\bar \xi_{2n})\rangle_{R_1} d^2\tilde \xi_1...d^2\tilde \xi_N,
 \end{eqnarray}
 where $C_n$ is the Jacobian  from $O(\omega_1)...O({\omega_n})$.
We know that when $L+l<t$ or $t<L$, $\xi_i-\xi_{i+1} \sim O(\epsilon)$ ($i=1,...,2n-1$). Then the leading order term in the $2n$-point correlation function
is
\begin{eqnarray}\label{2npointperturbation}
 &&\langle O^{\dag}(\omega_1,\bar \omega_1)O(\omega_2,\bar \omega_2)...O(\omega_{2n},\bar \omega_{2n})\rangle_{\Sigma_n}\nonumber \\
 &&\simeq \sum_{N=0}^{+\infty}C_n \frac{(-\lambda)^N}{N!}\int_{R_1}...\int_{R_1} (\omega'(\tilde \xi_1)...\omega'(\tilde \xi_N))^{2-\Delta} \langle \Phi(\tilde \xi_1)... \Phi(\tilde \xi_N)\rangle_{{{R_1}}} d^2\tilde \xi_1...d^2\tilde \xi_N\nonumber \\
 &&\prod_{i=1}^{n-1}\langle O^{\dag}(\xi_{2i+1},\bar \xi_{2i+1})O(\xi_{2i+2},\bar \xi_{2i+2})\rangle_{{R_1}}.
 \end{eqnarray}

 In (\ref{2npointperturbation}), the contribution from $O(z_i)$ being contracted with $\Phi(\bar \xi_1)$ is subleading in the limit $\ep\rightarrow 0$. There are some subtle issues which have been come across in perturbative CFT. It is well known that such integrals are potentially ambiguous due to the singularities from the contact terms when $O(z_i)$ contracts with $\Phi(\bar \xi_j)$\cite{Douglas:1993wy}\cite{Dijkgraaf:1996iy} . After proper regularization which highly depends on the specific choice of $O$ and $\Phi$, the contraction between $O$ and $\Phi$ make a finite contribution. That means the dominant contribution in the limit $\ep\rightarrow 0$ comes only from the contraction between $O$'s as shown in the last line of (\ref{2npointperturbation}). In order to present the crucial point clearly, we take the second R\'enyi entropy as an example.
 The second R\'enyi entropy (\ref{replica}) can be expanded by the powers of $\lambda$ as following
 \bea \label{example}
\Delta S_A^{(2)}&=& \log \Big[{\langle OOOO\rangle_{R_2}\over \langle OO\rangle_{R_1}^2}-2\lambda { \langle OOOO\rangle_{R_2}\over \langle OO\rangle_{R_1}}\int \langle\Phi(\tilde \xi) OO\rangle_{R_1} (\omega'(\tilde \xi))^{2-\Delta} d\tilde \xi\nonumber \\
&+&{\lambda C_2 \over \langle OO\rangle_{R_1}^2}\int d\tilde \xi(\omega'(\tilde \xi))^{2-\Delta}\langle\Phi(\tilde \xi) OOOO\rangle_{R_1}+...\Big],
 \eea
where the $...$ stands for the higher order terms in $g$. In order to make the perturbation well defined, we need a regularization to deal with the divergent terms   in the integration. Especially for $\int_{R_1}(\omega'(\tilde \xi))^{2-\Delta} \langle\Phi(\tilde \xi) O(w_1)O(w_2)\rangle_{R_1} d\tilde \xi$, we should introduce two contract terms like $\int (\omega'(\tilde \xi))^{2-\Delta}\langle\Phi(\tilde \xi) O(w_1)O(w_2)\rangle_{R_1} (\delta({\tilde \xi}- w_1)+ \delta({\tilde \xi}- w_2)) d\tilde \xi$ to subtract the divergence if $\Phi$ is close to $O$. For the second integration in (\ref{example}), we should do a similar regularization to deal with the divergence. All the divergent terms can be absorbed into $\lambda$ to make the perturbation well defined. Finally,  $\langle\Phi O(z_1)O(z_2)O(z_3)O(z_4)\rangle_{R_2}$ can be expressed by $\langle\Phi \rangle\langle O(z_1)O(z_2)O(z_3)O(z_4)\rangle_{R_1}$ in the early time limit or the late time limit (\ref{earlytimelate}). Repeating the analysis, we can calculate the $\Delta S_A^{(2)}$ to higher order of $\lambda$ formally. For example,
\bea \label{example1}
\Delta S_A^{(2)}&=& \log\Big[{\langle OOOO\rangle_{R_2}\over \langle OO\rangle_{R_1}^2}-2\lambda C_2{ \langle OOOO\rangle_{R_2}}\int \langle\Phi(\tilde \xi)\rangle_{R_1}(\omega'(\tilde\xi))^{2-\Delta} d\tilde \xi\nonumber \\
&+&{\lambda C_2 \langle OOOO\rangle_{R_2} \over \langle OO\rangle_{R_1}^2}\int d\tilde \xi (\omega'(\tilde \xi))^{2-\Delta}\langle\Phi(\tilde \xi)\rangle_{R_1}+...\Big]. \eea
In the relation (\ref{example1}), ${\langle OOOO\rangle_{R_2}\over \langle OO\rangle_{R_1}^2}=1/d_{O}$ which corresponds to the contribution from the local excitation during the time $l<t<L+l$.

If $\Phi$ is a primary field, $\langle \Phi \rangle_{R_1}=0$. We consider the second order of $\lambda$. The two-point function is
 \begin{eqnarray}
 \langle O^{\dag}(\omega_1,\bar \omega_1)O(\omega_2,\bar \omega_2)_{\Sigma_1}=\langle O^{\dag}(\omega_1,\bar \omega_1)O(\omega_2,\bar \omega_2)_{R_1}(1+\frac{1}{2}\lambda^2\int_{R_1} \langle \Phi(z_1) \Phi(z_2)\rangle_{R_1} d^2z_1 d^2 z_2).
 \end{eqnarray}
 Similarly, the $2n$-point function
 \begin{eqnarray}
 &&\langle O^{\dag}(\omega_1,\bar \omega_1)O(\omega_2,\bar \omega_2)...O(\omega_{2n},\bar \omega_{2n})\rangle_{\Sigma_n}\nonumber \\ \nonumber
 &&\simeq C_n \prod_{i=1}^{n-1} \langle O^{\dag}(\xi_{2i+1},\bar \xi_{2i+1})O(\xi_{2i+2},\bar \xi_{2i+2})\rangle (1+\frac{1}{2}\lambda^2 \int_{R_n} \langle \Phi(z_1) \Phi(z_2)\rangle_{R_n} d^2z_1 d^2 z_2),
 \end{eqnarray}
 during the time $L+l<t$ or $0<t<l$.
Then
\begin{eqnarray}\label{Pertubrationresult}
 \Delta S^{(n)}&=&\Delta S_{CFT}^{(n)} +\frac{1}{1-n}\log \frac{1+\frac{1}{2}\lambda^2 \int \langle \Phi(z_1)\Phi(z_2)\rangle_{R_n} d^2z_1d^2z_2 }{(1+\frac{1}{2}\lambda^2 \int \langle \Phi(z_1)\Phi(z_2)\rangle_{R_1} d^2z_1d^2z_2)^n}\nonumber \\
 &\simeq&\Delta S_{CFT}^{(n)} +\frac{1}{2(1-n)}\lambda^2\Big( \int \langle \Phi(z_1) \Phi(z_2)\rangle_{R_n} d^2z_1 d^2 z_2-n\int \langle \Phi(z_1) \Phi(z_2)\rangle_{R_1}d^2z_1d^2z_2\Big) ,\nonumber\\
 \end{eqnarray}
 where $\Delta S^{(n)}_{CFT}$ denotes the result in the CFT. The second part in (\ref{Pertubrationresult}) is just the correction to the R\'enyi entropy from the deformation. It is called global contribution in
 the paper \cite{Cardy:2010zs} . During the time $l<t<L+l$ the only change is in $\Delta S_{CFT}^{(n)}$, which contributes the $\log d_{O}$.

\section{Conclusion and Discussion}

In this paper, we have studied the R\'enyi entropy of local descendent operators in 2D CFT, extending the previous studies  in \cite{Nozaki:2014hna}\cite{He:2014mwa}\cite{Guo:2015uwa}.
In \cite{Nozaki:2014hna}\cite{He:2014mwa}\cite{Guo:2015uwa}, it has been found that the  R\'enyi entropy of a primary state is equal to the logarithmic of quantum dimension of the primary operator. It is a natural question to consider the quantum entanglement of the descendent states. Firstly, we showed that for the specific operator  $L_{-1} O$ with $O$ being primary, its  R\'enyi entropy  is still the logarithmic of quantum dimension of the primary operator $O$.  Secondly, we discuss the  descendent state $L^{(-)} O|0\rangle$ generated by a quasi-primary operator. For such quasi-primary states, we showed that their quantum entanglements are the same as their primaries.
Despite the fact that the operators look quite complicated and their conformal transformations are involved, the leading divergent terms in the early time and late time limit
are simple, behaving as the one for primary operators. As a result, the quantum entanglement of the quasi-primary operators are the same as the primaries. Moreover we discussed the most generic descendent operators of the form
\be V=\sum_{m,j,r,k}d_{m,j;r,k}(\partial^{m}L^{(-,j)})(\bar{\partial}^{r}\bar{L}^{(-,k)} )O_a(w,\bar{w}) .\ee
Out of surprise, we found that the R\'enyi entropy of such operator is generally different from the one of the primary $O_a(w,\bar{w})$. Only when the rank of the matrix $d_{m,j;r,k}$ is one, the entropies are the same as the ones of the primary. A typical example of such operator is
\be
V^{(1)}(w,\bar w)=\cal{L}^{(-)}\bar{\cal{L}}^{(-)}O.
\ee
Otherwise there is extra contribution. A typical example with extra contribution is  the operator of the form
\be
V^{(2)}(w,\bar w)=(L^{-}+\bar L^{-} )O(w,\bar w). \ee

To clarify the entropy difference between two kinds of operators $ V^{(1)}(w,\bar w)$ and $
V^{(2)}(w,\bar w)$, it would be  illuminating to consider the free scalar field theory $\phi$ in 2D.  Consider the primary operators $O_1$ and $O_2$, with
\begin{eqnarray}
O_1= e^{i\phi},\ \ O_2=\frac{1}{\sqrt{2}}(e^{i\phi}+e^{-i\phi}),
\end{eqnarray}
where $\phi\equiv \phi(z)+\bar \phi(\bar z)$. Following \cite{Nozaki:2014hna}\cite{He:2014mwa} the excited state $O_1\Ket{0}$ is regarded as the product state
$e^{i\phi(z)}\Ket{0}_L\otimes e^{i\bar \phi(\bar z)}\Ket{0}_R $ in the chiral and anti-chiral sectors. It is not an entangled state, so we get a vanishing entanglement entropy. On
the other hand, the operator $O_2$ creates the maximally entangled state, $\frac{1}{\sqrt{2}}(e^{i\phi(z)}\Ket{0}_L\otimes e^{i\bar \phi(\bar z)}\Ket{0}_R+e^{-i\phi(z)}\Ket{0}_L\otimes e^{-i\bar \phi(\bar z)}\Ket{0}_R)$. The R\'enyi entropy is $\log 2$ when one of the sector spreads into the region of
the subsystem.
The first descendent operators $O^{(-1)}_1=L_{-1} O_1$ and $O^{(-1)}_{2}=L_{-1} O_2$ are
\begin{eqnarray}
O_1^{(-1)}=i\partial \phi(z)  e^{i\phi},\ \ O_2^{(-1)}=\frac{i}{\sqrt{2}}\partial \phi(z)(e^{i\phi}+e^{-i\phi}).
\end{eqnarray}
The state $O_1^{(-1)} \Ket{0}=(i\partial\phi(z)e^{i\phi(z)})\Ket{0}_L\otimes e^{i\bar \phi(\bar z)}\Ket{0}_R$ is still a product state, has vanishing the R\'enyi entropy. Similarly, the state $O_2^{(-1)} \Ket{0}= \frac{1}{\sqrt{2}}[(i\partial\phi(z)e^{i\phi(z)})\Ket{0}_L\otimes e^{i\bar \phi(\bar z)}\Ket{0}_R-(i\partial\phi(z)e^{-i\phi(z)})\Ket{0}_L\otimes e^{-i\bar \phi(\bar z)}\Ket{0}_R]$ is still a maximally entangled state. Therefore, the R\'enyi entropy should be same as the original state. We could do the operation  on the chiral sector by $\prod_{i} L_{-i}^{k_i}$ ($i,k_i=1,2,...$) any times. The result is just a new EPR state. The
operation by $L_{-i}$ does not change the entanglement property, but just change the state in the chiral sector. This is the basic reason that the local excitation by the descendent operators works similarly  as the primary operator in this class. The same argument could also be used for the operations $\prod_{i} \bar L_{-i}^{k_i}$ ($i,k_i=1,2,...$).

Now Let us see what will happen when we apply the operation $\sum_{\{n_i\}\{n_j\}}(\prod_{i} L_{-n_i}+\prod_{j}{\bar L}_{-n_j})$, e.g., $L_{-1}+\bar L_{-1}$. The operators $O_1$ and $O_2$
become\footnote{We ignore the constant $\frac{1}{\sqrt{2}}$ for $O_2$ below.}
\begin{eqnarray}
O^{(-1,-\bar 1)}_1\equiv (L_{-1}+\bar L_{-1}) O_1=(i\partial \phi(z)+i\bar \partial \bar \phi(\bar z))e^{i\phi(z)+i\bar \phi(\bar z)},\nonumber \\
O^{(-1,-\bar 1)}_2 \equiv (L_{-1}+\bar L_{-1}) O_2=(i\partial \phi(z)+i\bar \partial \bar \phi(\bar z))(e^{i\phi}-e^{-i\phi}).
\end{eqnarray}
For the operator $O_1^{(-1,-\bar 1)}$, the corresponding state is
\begin{eqnarray}
O_1^{(-1,-\bar 1)}\Ket{0}=\Ket{2}_L\otimes \Ket{1}_R+\Ket{1}_L\otimes \Ket{2}_R,\nn
\end{eqnarray}
where we define
\begin{eqnarray}
\Ket{1}_{L}=e^{i\phi(z)}\Ket{0}_{L},\ \Ket{2}_L=i\partial \phi(z) e^{i\phi(z)}\Ket{0}_{L},\ \Ket{1}_{R}=e^{i\bar \phi(\bar z)}\Ket{0}_{R},\  \Ket{2}_R=i\bar \partial \bar \phi(\bar z) e^{i\bar \phi(\bar z)}\Ket{0}_{R}.
\end{eqnarray}
With some normalization this is just the EPR state, the entanglement entropy of which is $\log 2$ when one of the sector state spreads into the subsystem $A$.
In our calculation this contribution comes from the second term in (\ref{DecendentResult}). For the operator $O_2^{(-1,-\bar 1)}$, the state is
\begin{eqnarray}\label{O2EPR}
O_2^{(-1,-\bar 1)}\Ket{0}=(\Ket{2}_L\Ket{1}_R+\Ket{1}_L\Ket{2}_R)-(\Ket{4}_L\Ket{3}_R+\Ket{3}_L\Ket{4}_R),
\end{eqnarray}
where we define
\begin{eqnarray}
\Ket{3}_L=e^{-i\phi(z)}\Ket{0}_L,\ \Ket{4}_L=i\partial \phi(z)e^{-i\partial\phi(z)}\Ket{0}_L,\ \Ket{3}_R=e^{-i\bar \phi(\bar z)}\Ket{0}_R,\ \Ket{4}_R=i\bar \partial \bar \phi(\bar z)e^{-i\bar \partial \bar \phi(\bar z)}\Ket{0}_R.\nn
\end{eqnarray}
We could see that after  normalization the relation  (\ref{O2EPR}) is just the direct sum of two EPR states. The entanglement entropy is $2\log 2$ when one of the sector state spreads into the subsystem $A$. This is also consistent with our calculation (\ref{DecendentResult}).

Indeed for the interacting theory it is not easy to find such a clear explanation. Actually we still need better understanding on the fact that the seemingly decoupled chiral and anti-chiral sector are entangled with each other, even for the primary excitation state, see \cite{Jackson:2014nla}.

In this paper,  we also introduced a boundary which do not break conformal symmetry as in \cite{Guo:2015uwa}. For the descendent state of type (\ref{first-eq}), we estimated  the maximal value of R\'enyi entropy of local descendent states, which is the same as the one in the theory without such kind of boundary, although the boundary change the time evolution behavior of R\'enyi entropy. Finally, we  deformed the original CFT by additional operator and discussed the R\'enyi entropy of generic local excited states. If the deformation is small, we can treat it as a perturbation of CFT. By this way, we can estimate that this deformation make a global contribution to the R\'enyi entropy up to $g^2$, if the deformation is a primary field. Though there are divergences in the global contribution\cite{Cardy:2010zs}, such divergences can be regulated appropriately. Actually there are various definite extensive examples \cite{Gaberdiel:2013jca}\cite{Datta:2014ska}\cite{Datta:2014uxa}\cite{Long:2014oxa}\cite{Datta:2014zpa} to show that global contribution is finite.


\vspace*{10mm}
\noindent {\large{\bf Acknowledgments}}\\

We are grateful to J. L. Cardy, Mitsutoshi Fujita, Rene Meyer, Masahiro Nozaki, T.~Numasawa, Tadashi Takayanagi and K.~Watanabe for useful conversations and correspondence. W.G. and S.H. thank Miao Li, Tadashi Takayanagi for their encouragement and support. S. H would like to thank {to} Sichuan University, IPMU, and Max Planck Institute for Gravitational Physics (AEI) for hospitality. BC and J.q. Wu was supported in part by NSFC Grants No. 11275010, No. 11335012 and No. 11325522. W. Z. Guo is supported by Postgraduate Scholarship Program of China Scholarship Council. S.H. is supported by JSPS postdoctoral fellowship for foreign researchers and by the National Natural Science Foundation of China (No.11305235).
\vspace*{5mm}

\begin{appendix}
\section{Conformal Transformation for Descendent Operator}

In this section, we show how to take the conformal transformation for a decedent operator. For a general primary operator, under a conformal transformation it acts as
\be \psi(z)=\psi(w)(\frac{\partial w}{\partial z})^h(\frac{\partial\bar{w}}{\partial \bar{z}})^{\bar{h}}. \ee
First consider the conformal transformation of $L_{-1}\psi(z)$
\bea L_{-1}\psi(z)&=&\frac{1}{2\pi i}\oint dz^{'} z^{'n+1}T(z^{'})\psi(z) \nn\\
&=&\frac{1}{2\pi i}\oint dz^{'}[T(w^{'})[\frac{\partial w^{'}}{\partial z^{'}}+\frac{c}{12}\{w^{'},z^{'}\}]
\psi(w)(\frac{\partial w}{\partial z})^h(\frac{\partial \bar{w}}{\partial \bar{z}})^{\bar{h}} \nn\\
&=&\frac{1}{2\pi i}\oint dz^{'}[T(w^{'})(\frac{\partial w^{'}}{\partial z^{'}})^2]\psi(w)
(\frac{\partial w}{\partial z})^h(\frac{\partial \bar{w}}{\partial \bar{z}})^{\bar{h}} \nn\\
&=&\frac{1}{2\pi i}\oint dz^{'}(\frac{\partial w^{'}}{\partial z^{'}})^2[\frac{h\psi(w)}{(w^{'}-w)^2}
+\frac{\partial_w\psi(w)}{w^{'}-w}+\mbox{normal}]
(\frac{\partial w}{\partial z})^h(\frac{\partial \bar{w}}{\partial \bar{z}})^{\bar{h}} \nn\\
&=&(\frac{\frac{\partial^2 w}{\partial z^2}}{\frac{\partial w}{\partial z}}h\psi(w)
+(\frac{\partial w}{\partial z})\partial_w\psi(w))
(\frac{\partial w}{\partial z})^h(\frac{\partial \bar{w}}{\partial \bar{z}})^{\bar{h}}
\eea
This is really the conformal transformation for the operator $\partial \psi(z)$, which support our calculation.

For more generic descendent operator
\bea L_{-n}\psi (z)&=& \frac{1}{2\pi i}\oint dz^{'}(z^{'}-z)^{-n+1}T(z^{'})\psi(z) \nn\\
&=&\frac{1}{2\pi i}\oint dz^{'}(z^{'}-z)^{-n+1}
[T(w^{'})(\frac {\partial w^{'}}{\partial z^{'}})^2+\frac{c}{12}\{w^{'},z^{'}\}]\psi(w)
(\frac{\partial w}{\partial z})^h(\frac{\partial \bar{w}}{\partial \bar{z}})^{\bar{h}} \nn\\
&=&\frac{1}{2\pi i}\oint dz^{'}(z^{'}-z)^{-n+1}
[(\frac{\partial w^{'}}{\partial z^{'}})^2(
\frac{h\psi(w)}{(w^{'}-w)^2}+\frac{\partial_w\psi(w)}{w^{'}-w}+:T(w^{'})\psi(w):)\nn\\&~&
+\frac{c}{12}\{w^{'},z^{'}\}\psi(w)](\frac{\partial w}{\partial z})^h
(\frac{\partial \bar{w}}{\partial \bar{z}})^{\bar{h}}
\eea
Calculating the contour integral, the conformal transformation of specific descendent operator is
\bea L_{-2}\psi(z)&=&
(-\frac{1}{4}(\frac{\frac{\partial^2 w}{\partial z^2}}{\frac{\partial w}{\partial z}})^2
+\frac{2}{3}\frac{\frac{\partial^3w}{\partial z^3}}{\frac{\partial w}{\partial z}})
h\psi(w)(\frac{\partial w}{\partial z})^h(\frac{\partial \bar{w}}{\partial \bar{z}})^{\bar{h}} \nn\\
&+&\frac{3}{2}\frac{\partial^2 w}{\partial z^2} \partial_w \psi(w)
(\frac{\partial w}{\partial z})^h(\frac{\partial \bar{w}}{\partial \bar{z}})^{\bar{h}}
+(\frac{\partial w}{\partial z})^2L_{-2}\psi(w)(\frac{\partial w}{\partial z})^h
(\frac{\partial \bar{w}}{\partial \bar{z}})^{\bar{h}}\nn \\
&~&+\frac{c}{12}\{w,z\}(\frac{\partial w}{\partial z})^2\psi(w)
(\frac{\partial w}{\partial z})^h(\frac{\partial \bar{w}}{\partial \bar{z}})^{\bar{h}}
\eea

Actually, we can get a more general conformal transformation for the operators, which is an expansion for the previous result
\bea L_{-2}\mid_{z_1}&=&\frac{1}{2\pi i}\oint dz \frac{1}{z-z_1}T(z) \nn\\
&=&\frac{1}{2\pi i}\oint dz\frac{1}{z-z_1}(T(w)(\frac{\partial w}{\partial z})^2+\frac{c}{12}\{w,z\}) \nn\\
&=&\frac{1}{2\pi i}\oint dz \frac{1}{z-z_1}((w-w_1)L_{-3}+L_{-2}+\frac{L_{-1}}{w-w_1}+\frac{L_0}{(w-w_1)^2}
+\frac{L_1}{(w-w_1)^3}+\frac{L_2}{(w-w_1)^4} \notag \\
&~& +\mbox{higher order})(\frac{\partial w}{\partial z})^2
+\frac{c}{12}\{w_1,z_1\} \notag \\
&=&L_{-2}(\frac{\partial w_1}{\partial z_1})^2+L_{-1}\frac{3}{2}\frac{\partial^2 w_1}{\partial z_1^2}
+L_0(-\frac{1}{4}(\frac{\frac{\partial^2 w_1}{\partial z_1^2}}{\frac{\partial w_1}{\partial z_1}})^2
+\frac{2}{3}\frac{\frac{\partial^3 w_1}{\partial z_1^3}}{\frac{\partial w_1}{\partial z_1}}) \nn\\
&~&+L_1\frac{1}{\frac{\partial w_1}{\partial z_1}}\frac{1}{24}
(6(\frac{\frac{\partial^2 w_1}{\partial z_1^2}}{\frac{\partial w_1}{\partial z_1}})^3
-12\frac{\frac{\partial^2 w_1}{\partial z_1^2}}{\frac{\partial^3 w_1}{\partial z_1^3}}
+5\frac{\frac{\partial^4 w_1}{\partial z_1^4}}{\frac{\partial w_1}{\partial z_1}})\nn \\
&~&+L_2\frac{1}{(\frac{\partial w_1}{\partial z_1})^2}\frac{1}{720}
[-225(\frac{\frac{\partial^2 w_1}{\partial z_1^2}}{\frac{\partial w_1}{\partial z_1}})^4
+480(\frac{\frac{\partial^2 w_1}{\partial z_1^2}}{\frac{\partial w_1}{\partial z_1}})^2
\frac{\frac{\partial^3 w_1}{\partial z_1^3}}{\frac{\partial w_1}{\partial z_1}}
-100(\frac{\frac{\partial ^3w_1}{\partial z_1^3}}{\frac{\partial w_1}{\partial z_1}})^2
-180\frac{\frac{\partial ^2w_1}{\partial z_1^2}}{\frac{\partial w_1}{\partial z_1}}
\frac{\frac{\partial ^4w_1}{\partial z_1^4}}{\frac{\partial w_1}{\partial z_1}} \nn\\
&~&+36\frac{\frac{\partial^5 w_1}{\partial z_1^5}}{\frac{\partial w_1}{\partial z_1}}]
+\mbox{higher orders}
\eea

\end{appendix}




\vspace*{5mm}
\begin{thebibliography}{99}





  \bibitem{wen}
  A.~Kitaev and J.~Preskill,
  ``Topological entanglement entropy,''
  Phys.\ Rev.\ Lett.\  {\bf 96}, 110404 (2006)
  [hep-th/0510092].
M.~Levin and X.~G.~Wen,
''Detecting topological order in a ground state wave function,''
  Phys.\ Rev.\ Lett.\  {\bf 96}, 110405 (2006)
 [arXiv:cond-mat/0510613].


\bibitem{FFN}
 P.~Fendley, M.~P.~A.~Fisher and C.~Nayak,
 ``Topological Entanglement Entropy from the Holographic Partition Function,''
 J.\ Statist.\ Phys.\  {\bf 126} (2007) 1111.  




\bibitem{CC}
  P.~Calabrese and J.~L.~Cardy,
  ``Entanglement entropy and quantum field theory,''
  J.\ Stat.\ Mech.\  {\bf 0406}, P06002 (2004)
  [hep-th/0405152].

\bibitem{Moore:1988uz}
  G.~W.~Moore and N.~Seiberg,
 { \it Polynomial Equations for Rational Conformal Field Theories,}
  Phys.\ Lett.\ B {\bf 212}, 451 (1988);
  {\it Classical and Quantum Conformal Field Theory,}
  Commun.\ Math.\ Phys.\  {\bf 123}, 177 (1989).

\bibitem{Verlinde:1988sn}
  E.~P.~Verlinde,
  {\it Fusion Rules and Modular Transformations in 2D Conformal Field Theory,''}
  Nucl.\ Phys.\ B {\bf 300}, 360 (1988);  J.~L.~Cardy,
  {\it Boundary Conditions, Fusion Rules and the Verlinde Formula,''}
  Nucl.\ Phys.\ B {\bf 324}, 581 (1989);  R.~Dijkgraaf and E.~P.~Verlinde,
 {\it Modular Invariance And The Fusion Algebra,}
  Nucl.\ Phys.\ Proc.\ Suppl.\  {\bf 5B}, 87 (1988).


\bibitem{Alcaraz:2011tn}
  F.~C.~Alcaraz, M.~I.~Berganza and G.~Sierra,
  ``Entanglement of low-energy excitations in Conformal Field Theory,''
  Phys.\ Rev.\ Lett.\  {\bf 106}, 201601 (2011)
  [arXiv:1101.2881 [cond-mat.stat-mech]].

\bibitem{Palmai:2014jqa}
  T.~Pálmai,
  ``Excited state entanglement in one dimensional quantum critical systems: Extensivity and the role of microscopic details,''
  Phys.\ Rev.\ B {\bf 90}, no. 16, 161404 (2014)
  [arXiv:1406.3182 [hep-th]].








\bibitem{Nozaki:2014hna}
  M.~Nozaki, T.~Numasawa and T.~Takayanagi,
  ``Quantum Entanglement of Local Operators in Conformal Field Theories,''
  Phys.\ Rev.\ Lett.\  {\bf 112}, 111602 (2014)
  [arXiv:1401.0539 [hep-th]].


\bibitem{Nozaki:2014uaa}
  M.~Nozaki,
  ``Notes on Quantum Entanglement of Local Operators,''
  JHEP {\bf 1410}, 147 (2014)
  [arXiv:1405.5875 [hep-th]].

\bibitem{Shiba:2014uia}
  N.~Shiba,
  ``Entanglement Entropy of Disjoint Regions in Excited States : An Operator Method,''
  JHEP {\bf 1412} (2014) 152
  [arXiv:1408.0637 [hep-th]].

\bibitem{He:2014mwa}
  S.~He, T.~Numasawa, T.~Takayanagi and K.~Watanabe,
  ``Quantum Dimension as Entanglement Entropy in 2D CFTs,''
  Phys.\ Rev.\ D {\bf 90}, 041701 (2014)
  [arXiv:1403.0702 [hep-th]].

\bibitem{Caputa:2014vaa}
  P.~Caputa, M.~Nozaki and T.~Takayanagi,
  ``Entanglement of local operators in large-N conformal field theories,''
  PTEP {\bf 2014}, 093B06 (2014)
  [arXiv:1405.5946 [hep-th]].

\bibitem{Asplund:2014coa}
  C.~T.~Asplund, A.~Bernamonti, F.~Galli and T.~Hartman,
  ``Holographic Entanglement Entropy from 2d CFT: Heavy States and Local Quenches,''
  JHEP {\bf 1502}, 171 (2015)
  [arXiv:1410.1392 [hep-th]].

\bibitem{Caputa:2014eta}
  P.~Caputa, J.~Simón, A.~¦tikonas and T.~Takayanagi,
  JHEP {\bf 1501}, 102 (2015)
  [arXiv:1410.2287 [hep-th]].

\bibitem{Guo:2015uwa}
  W.~Z.~Guo and S.~He,
  ``R\'enyi entropy of locally excited states with thermal and boundary effect in 2D CFTs,''
  JHEP {\bf 1504}, 099 (2015)
  [arXiv:1501.00757 [hep-th]].

  \bibitem{Das:2015oha}
  D.~Das and S.~Datta,
  ``Universal features of left-right entanglement entropy,''
  arXiv:1504.02475 [hep-th].




\bibitem{Calabrese:2005in}
  P.~Calabrese and J.~L.~Cardy,
  ``Evolution of entanglement entropy in one-dimensional systems,''
  J.\ Stat.\ Mech.\  {\bf 0504}, P04010 (2005)
  [cond-mat/0503393].

\bibitem{Dotsenko:1984nm}
  V.~S.~Dotsenko and V.~A.~Fateev,
  ``Conformal Algebra and Multipoint Correlation Functions in Two-Dimensional Statistical Models,''
  Nucl.\ Phys.\ B {\bf 240}, 312 (1984).

\bibitem{Dotsenko:1984ad}
  V.~S.~Dotsenko and V.~A.~Fateev,
  ``Four Point Correlation Functions and the Operator Algebra in the Two-Dimensional Conformal Invariant Theories with the Central Charge $c < 1$,''
  Nucl.\ Phys.\ B {\bf 251}, 691 (1985).

  \bibitem{CFT}
Ph. Di Francesco, P. Mathieu, D. S\'en\'echal, ``Conformal Field Theory,'' Springer 1998.



  \bibitem{BPZ}
A.~Belavin, A.~M. Polyakov, and A.~Zamolodchikov, {\it {Infinite Conformal
  Symmetry in Two-Dimensional Quantum Field Theory}},  {\em Nucl.Phys.} {\bf
  B241} (1984) 333--380;
S.~Ferrara, A.~Grillo, and R.~Gatto, {\it {Tensor representations of conformal
  algebra and conformally covariant operator product expansion}},  {\em Annals
  Phys.} {\bf 76} (1973) 161--188;
A.~M. Polyakov, {\it {Non-Hamiltonian Approach to the Quantum Field Theory at
  Small Distances}},  {\em Zh.Eksp.Teor.Fiz.} (1974).

\bibitem{Osborn:2012vt}
  H.~Osborn,
  ``Conformal Blocks for Arbitrary Spins in Two Dimensions,''
  Phys.\ Lett.\ B {\bf 718}, 169 (2012)
  [arXiv:1205.1941 [hep-th]].

\bibitem{Cardy1}
John L. Cardy,  "Conformal invariance and surface critical behavior." Nuclear Physics B 240.4 (1984): 514-532.

\bibitem{Cardy2}
John L. Cardy,  "Boundary conditions, fusion rules and the Verlinde formula." Nuclear Physics B 324.3 (1989): 581-596.

\bibitem{Cardy:2010zs}
  J.~Cardy and P.~Calabrese,
  ``Unusual Corrections to Scaling in Entanglement Entropy,''
  J.\ Stat.\ Mech.\  {\bf 1004}, P04023 (2010)
  [arXiv:1002.4353 [cond-mat.stat-mech]].

\bibitem{Douglas:1993wy}
  M.~R.~Douglas,
  ``Conformal field theory techniques in large N Yang-Mills theory,''
  hep-th/9311130.

\bibitem{Dijkgraaf:1996iy}
  R.~Dijkgraaf,
  ``Chiral deformations of conformal field theories,''
  Nucl.\ Phys.\ B {\bf 493}, 588 (1997)
  [hep-th/9609022].


\bibitem{Gaberdiel:2013jca}
  M.~R.~Gaberdiel, K.~Jin and E.~Perlmutter,
  ``Probing higher spin black holes from CFT,''
  JHEP {\bf 1310}, 045 (2013)
  [arXiv:1307.2221 [hep-th]].

\bibitem{Datta:2014ska}
  S.~Datta, J.~R.~David, M.~Ferlaino and S.~P.~Kumar,
  ``Higher spin entanglement entropy from CFT,''
  JHEP {\bf 1406}, 096 (2014)
  [arXiv:1402.0007 [hep-th]].

\bibitem{Datta:2014uxa}
  S.~Datta, J.~R.~David, M.~Ferlaino and S.~P.~Kumar,
  ``Universal correction to higher spin entanglement entropy,''
  Phys.\ Rev.\ D {\bf 90}, no. 4, 041903 (2014)
  [arXiv:1405.0015 [hep-th]].

\bibitem{Long:2014oxa}
  J.~Long,
  ``Higher Spin Entanglement Entropy,''
  JHEP {\bf 1412}, 055 (2014)
  [arXiv:1408.1298 [hep-th]].

\bibitem{Datta:2014zpa}
  S.~Datta, J.~R.~David and S.~P.~Kumar,
  ``Conformal perturbation theory and higher spin entanglement entropy on the torus,''
  JHEP {\bf 1504}, 041 (2015)
  [arXiv:1412.3946 [hep-th]].

\bibitem{Jackson:2014nla}
  S.~Jackson, L.~McGough and H.~Verlinde,
  ``Conformal Bootstrap, Universality and Gravitational Scattering,''
  arXiv:1412.5205 [hep-th].

\bibitem{Pawel}
  Pawel Caputa, Alvaro Veliz-Osorio,
  ``Entanglement constant for conformal families,''
  arXiv:1507.00582[hep-th].

\end{thebibliography}
\end{document}